\documentclass[onecolumn,showpacs]{revtex4}

\usepackage{graphicx}%
\usepackage{dcolumn}
\usepackage{amsmath}

\makeatletter
\def\btt#1{\texttt{\@backslashchar#1}}%
\DeclareRobustCommand\bblash{\btt{\@backslashchar}}%
\makeatother

\begin{document}

%\preprint{HEP/123-qed}

\title{Two-particle correlated interference in reflection: extending the quantum-classical boundary via a macroscopic quantum superposition insensitive to decoherence}
% Force line breaks with \\

\author{F.V. Kowalski and R.S. Browne}
\affiliation{Physics Department, Colorado
School of Mines, Golden CO. 80401 U.S.A.}

%\date{\today}% It is always \today, today, but you may specify any date with \date.
\begin{abstract}

Reflection of a microscopic particle from a mesoscopic/macroscopic `mirror' generates two-body correlated interference from the incident and reflected particle substates and their associated mirror substates. The microscopic momentum exchanged generates two mirror substates which interfere to produce fringes which do not vanish as the mirror mass increases. The small displacement between these mirror states can yield negligible environmental decoherence times. Mirror coherence lengths impose constraints on the extent of this interference, which are mitigated using interference of the two-body states associated with the particle reflecting from both of the two surfaces of a slab of matter in a manner analogous to the classical interference of a pulse of light reflecting from a `thin film'. This two-body correlated interference is modeled as a particle traversing a finite well with both the particle and well treated quantum mechanically. Such a treatment predicts the expected `thin-film' interference but only as a special case of a more general result. It is also shown that measurements on only the reflected particle (yielding a marginal probability density function) can act as a probe to reveal the quantum state of the macroscopic reflector. For equal masses, coherence of the particle substate is transferred to the mirror substate, a quantum manifestation of a familiar classical result.

\end{abstract}

\pacs{03.65.Ta, 03.65.-w,03.75.Dg,03.75.-b}

\maketitle
% ----------------------------------------------------------------

\section{Introduction}
\label{sec:intro}

Correlation and interference distinguish quantum from classical mechanics. The former is manifest in the measurement of many-body coincidences predicted by a quantum joint probability density function (PDF), which are not observed classically \cite{bell}. The latter is most familiar as a probability density function for an outcome that can be achieved in at least two indistinguishable ways. However, interference can also be generated by superposing many-body states in indistinguishable ways \cite{Gottfried}. Experimental confirmation of quantum correlation has involved photons \cite{Salart}, atoms \cite{Wineland}, and Josephson phase qubits \cite{Ansmann}. However, there is little experimental evidence of correlated interference between massive particles.

The boundary between the classical and quantum regimes continues to be an area of interest both in attempts to explain why quantum effects are difficult to measure on a macroscopic scale \cite{hagar} and in attempts to measure quantum effects on a macroscopic object \cite{Wineland,cronin,juffmann}. Reflection measurement methods used in an attempt to extend this boundary involve light interacting with stationary \cite{abbott} and oscillating mirrors \cite{aspelmeyer} and with microwaves interacting with an oscillating mirror \cite{oconnell}. Reflection of a single photon from an oscillating mirror \cite{nunnenkamp} and using a Bose Einstein condensate as a mirror from which photons reflect \cite{demartini} have been studied. 

To extend this boundary to larger masses, we model correlation of a mesoscopic/macroscopic `mirror' with a  non-zero rest mass microscopic `particle' after elastic reflection. Motion is in free space along one dimension and all states are unbound. Measurements of particle reflection, but not associated with correlated interference, have involved mirrors that reflect atoms \cite{kouznetsov} and Bose-Einstein condensates \cite{pasquini}, atoms reflecting from a solid surface \cite{shimizu}, neutrons \cite{hils} and atoms \cite{colombe} reflecting from vibrating mirrors, and atoms reflecting from a switchable mirror \cite{szriftgiser}.

If the particle and mirror are initially in eigenstates of energy then the following two states interfere: the particle-mirror state before the particle reflects from the mirror and the particle-mirror state after reflection (interference is a consequence of the incident and reflected states being indistinguishable for a measurement of position). This is similar to the standing wave interference of a harmonic electromagnetic wave reflecting from a stationary mirror \cite{rauch}. Classically, however, the incident and reflected waves interfere while the mirror experiences only a continuous force due to radiation pressure.

The two-body quantum analogy involves solving the Schr\"odinger equation with an interaction potential modeling the reflection. Interference is expected between the incident and reflected particle substates {\em along with} interference of the mirror substates which have and have not reflected the particle. Their correlation is perhaps not expected, being a consequence of the solution to the Schr\"odinger equation from which a joint probability density function is constructed. This joint PDF then describes the correlations in the two-body interference which are manifest as coincidence rates, e.g. a correlation in the simultaneous measurement of particle {\em and} mirror positions. Asynchronous measurement in such a system has also been discussed \cite{kowalski}.

Any experimental realization of such an interferometer will involve wavegroups. While the coherence length for the particle currently can be as large as $10^{-6}$ m for atoms released from a Bose-Einstein condensate, that for a mesoscopic/macroscopic mirror will be many orders of magnitude smaller. However, the coherence length of the two-body state involves that of both the particle and mirror substates. The two-body interference in reflection from a mirror is then limited spatially (to the overlap region) and temporally (to the time of overlap) in a manner similar to that of a pulse of light reflecting from a moving mirror. 

Classical interference in reflection is more robust when the pulse of light reflects from a moving thin film for two main reasons: first the reflected pulses from each thin film surface maintain overlap as they both travel in the backward direction and second they have the same Doppler shift (since they reflect from interfaces moving at the same speed). Interference is then limited by a pulse coherence length which at its minimum is of order the film thickness. 

To achieve such robust two-body quantum interference we consider the analogous two-body situation, an example of which is a neutron reflecting from each of the two surfaces of a moving aluminum `slab.'  It is shown below that this two-body interference has the same `non-local' characteristics as does a pulse of light reflecting from a moving thin film. Treating both the particle and reflector as quantum objects yields the expected results of a one-body treatment (analogous to interference for a pulse of light reflecting from a moving thin film) but only as a special case. In addition, this treatment predicts novel reflection regimes depending on the coherence lengths of the particle and reflector. 

Quantum two-body reflection differs from that classically. For example, the classical kinematics, associated with reflection of a wavegroup, involves accelerated mirror motion while the kinematics associated with the expectation value of the position of either the particle or mirror, during the interaction, depends on the two-body wavegroup parameters such as coherence lengths. This is discussed in section \ref{sec:summary}. Also, under conditions of destructive interference, the particle and slab will never simultaneously be observed. Classically, only the pulse of light will never be observed. 

A simultaneous measurement, however, is difficult to perform. A much simpler method involves only measuring the reflected particle. Such a one-body measurement can reveal the quantum state of the reflector due to the effects of the reflector's coherence length on the two-body PDF, from which the one-body marginal PDF is derived. This manifestation of the macroscopic quantum state of the reflector on the interference measured only on the particle may be a practical way to verify that the mesoscopic/macroscopic reflector is indeed in a quantum state and thereby extend the quantum-classical boundary to larger masses.

Finding examples of mesoscopic/macroscopic quantum phenomena has been a topic of interest since early in the development of quantum mechanics, particularly with regard to validation issues. This continues to be a subject of interest. Consider next some of the arguments used to explain why such effects are not observed with mesoscopic/macroscopic objects, in relation to the two-body treatment of particle reflection: (1) In an introductory quantum course, the lack of such evidence is often justified via the double slit experiment, where the fringe spacing becomes imperceptible as the mass increases. In the comments after eqn. \ref{eq:fringespacing}, an explanation is given as to why the mirror fringe spacing does not shrink as its mass increases. (2) Another concern is if quantum correlation is indeed generated. The initial particle-mirror separable system becomes correlated in reflection as a consequence of conservation of energy and momentum. The discussion in section \ref{sec:summary} demonstrates why interferometric quantum correlation is a requisite in satisfying these conservation laws. (3) The difference in wavevectors of the interfering mesoscopic/macroscopic mirror substates, due to reflecting a microscopic particle, is much smaller than the spread in wavevectors of which the mirror wavegroup is composed, due to thermal motion. This effect is shown to not destroy interference in section \ref{sec:summary}. (4) More recently, decoherence of the macroscopic superposition via interaction with the environment has been used to explain the lack of interference. Mitigation of this and other decoherence mechanisms are discussed in section \ref{sec:decoher}. This work examines the assumptions of these arguments in the context of reflection of a microscopic particle from a mesoscopic/macroscopic mirror. 

A synopsis of this work is as follows: First the Schr\"odinger equation is solved for the two-body energy eigenstates in reflection from a moving mirror. Wavegroups formed from these solutions are then used to illustrate interference in reflection with an emphasis on variations in coherence lengths of the mirror. Coherence transfer from the particle to the mirror is also demonstrated and related to classical reflection. Two models are next developed for two-body reflection of a particle from a slab, whose classical analog is that of reflection of a pulse of light from a moving thin film. The effects of the coherence lengths of both the particle and slab are then discussed. Measurements of either the particle but not the slab and the slab but not the particle are then considered. Finally, decoherence issues, as well as methods to measure a mesoscopic/macroscopic reflector in a superposition state, are addressed.

The goal of this paper is to illustrate the unique interferometric properties of correlation interferometry in reflection. These properties are most easily understood using the example of a microscopic particle reflecting from a mesoscopic/macroscopic mirror. However, the focus on this example is not intended as a practical experimental proposal. Observation of such correlated interference will most likely first occur with a microscopic particle reflecting from a ``microscopic'' mirror.

\section{Particle-mirror interaction}
\label{sec:particlmirror}

\subsection{Two-body Schr\"odinger equation solutions}
\label{sec:theoryparticlemirror}

Before reflection, the solution to the Schr\"odinger equation for the non-interacting particle-mirror state is
\begin{eqnarray}
\Psi_{0} \propto \exp[i (k x_{1}-\frac{\hbar k^{2}}{2m}t+K x_{2}-\frac{\hbar K^{2}}{2M}t)],
\label{eq:PsiSeparable}
\end{eqnarray}
where $x_{1}$ and $x_{2}$ are the particle and mirror positions along the x-axis while $k$ and $K$ are the wavevectors for the particle and mirror respectively; $k=m v/\hbar$ and $K=M V/\hbar$ with masses $m$, $M$, and initial velocities $v$ and $V$, respectively. A wavegroup constructed from such uncorrelated particle-mirror states then leads to predictions about the probability of simultaneously finding the particle at $x_{1}$ and mirror at $x_{2}$.

The particle-mirror interaction is modeled as a moving delta function potential where reflection is assumed to occur at the center of mass (cm) of the mirror with the Schr\"odinger equation given by
\begin{eqnarray}
(\hbar^{2} \partial_{x_{1}}^{2}/2m+\hbar^{2} \partial_{x_{2}}^{2}/2M+\beta \delta[x_{1}-x_{2}] \notag
+i\partial_{t})\Psi=0.
\label{eq:Schreqn}
\end{eqnarray}
where square brackets are used to indicate the argument of a function. The mirror reflectivity, related to $\beta$, goes to infinity for a lossless mirror. The solution yields an energy eigenstates for the particle-mirror interaction (`harmonic wavefunctions'), for which neither the particle nor the mirror is localized.

A separable solution to the two-body Schr\"odinger equation results from a transformation to the center of mass (cm) and relative (rel) system (not to be confused with the cm of the particle or mirror). This does not change the total energy,  $E=(\hbar K)^{2}/(2M)+(\hbar k)^{2}/(2m)=E_{rel}+ E_{cm}$. The transformed Schr\"odinger equation becomes
\begin{eqnarray}
(\frac{\hbar^{2} \partial_{x_{cm}}^{2}}{2M_{tot}}+\frac{\hbar^{2} \partial_{x_{rel}}^{2}}{2 \mu}+\beta \delta[x_{rel}] \notag
+i\partial_{t})\Psi[x_{cm},x_{rel},t]=0
\label{eq:Scheqncm}
\end{eqnarray}
where $M_{tot}=m+M$, $\mu=mM/(m+M)$, $x_{cm}=(mx_{1}+Mx_{2})/M_{tot}$, and $x_{rel}=x_{1}-x_{2}$. Using 
\begin{eqnarray}
\Psi[x_{cm},x_{rel},t]=\psi_{cm}\psi_{rel}  \notag \\ =e^{-i E_{cm} t/\hbar} U[x_{cm}] e^{-i E_{rel} t/\hbar}u[x_{rel}]  \notag,
\label{eq:Schtot}
\end{eqnarray}
reduces the Schr\"odinger equation to two ordinary differential equations:
\begin{equation}
-\frac{\hbar^{2}}{2M_{tot}} \frac{d^{2}U[x_{cm}]}{dx_{cm}^{2}} = E_{cm} U[x_{cm}] 
\label{eq:ScheqODE1}
\end{equation}
\begin{equation}
-\frac{\hbar^{2}}{2 \mu} \frac{d^{2}u[x_{rel}]}{dx_{rel}^{2}}+\beta \delta[x_{rel}] = E_{rel} u[x_{rel}].
\label{eq:ScheqODE2}
\end{equation}

The particle-mirror solution must vanish at $x_{1}=x_{2}$ to satisfy the boundary condition at the mirror and not exist for $x_{rel}<0$ (or $x_{1}>x_{2}$) since the particle cannot move through the mirror (for the uncorrelated incident state, however, there is no interaction and the particle can move past the mirror).

In this transformed system, a solution is constructed from the superposition of incident and ``reflected'' wavefunctions,
\begin{equation}
\psi_{rel}=(e ^{i \phi_{in}}-e ^{i \phi_{ref}}) \theta [x_{rel}],
\label{eq:superposition}
\end{equation}
where $\theta [x_{rel}]$ is the unit step function. The only difference between the arguments of the two exponentials is the sign of the relative wavevector $K_{rel}$ which, due to reflection in the relative coordinate, is negative. That is,
\begin{eqnarray}
\phi_{in/ref}= \mbox{\boldmath$\pm$}~K_{rel} x_{rel}-\hbar K_{rel}^{2}t/2 \mu ,
\label{eq:ScheqPhase}
\end{eqnarray}
where the initial velocities must allow reflection to occur. 

The solution to eqn. \ref{eq:ScheqODE1} is given by 
\begin{equation}
\psi_{cm}=e ^{i (K_{cm} x_{cm}-\hbar K_{cm}^{2}t/2M_{tot}}).
\label{eq:Psicm}
\end{equation}
The complete solution is then $\Psi[x_{cm},x_{rel},t]=\psi_{cm}\psi_{rel}$. In this separable system the probability density is
\begin{eqnarray}
\Psi \Psi^{*}=4 \sin^{2}[K_{rel}x_{rel}].
\label{eq:SeparableInterference}
\end{eqnarray}

The cm-rel transformation does not give a solution for a physically realizable system. There exists no particle with a reduced mass, for example. Nevertheless, its utility lies in inverting the solution from cm and relative substates into particle and mirror substates. This yields a correlated particle-mirror state. An example of this procedure is found in the solution to the hydrogen atom where the Schr\"odinger equation is first transformed from the laboratory to the cm-relative coordinates yielding uncorrelated substates. However, expressing this result in the electron and proton substates reveals a correlation between them \cite{Tommasini}.

This change of partition is accomplished by the following substitutions in the separable solution given in eqn. \ref{eq:SeparableInterference}: $K_{cm}=k+K$, $K_{rel}=(Mk-mK)/M_{tot}$, $x_{rel}=x_{1}-x_{2}$, $x_{cm}=(mx_{1}+Mx_{2})/M_{tot}$, $E_{rel}=\hbar^2K_{rel}^{2}/2\mu$, and $E_{cm}=\hbar^2K_{cm}^{2}/2(m+M)$. These constraints transform $e ^{i \phi_{in}}$ in eqn. \ref{eq:superposition} into the non-interacting particle-mirror solution of eqn. \ref{eq:PsiSeparable}. Transformation of the energy component of each reflected substate, given by $p_{ref}^{2}/(2m)$ with $p_{ref}=\hbar \partial \phi_{ref}/\partial x_{1}$ for the reflected particle and $P_{ref}^{2}/(2M)$ with $P_{ref}=\hbar \partial \phi_{ref}/\partial x_{2}$ for the mirror, is consistent with that of reflection classically. 

With this change in coordinates eqn. \ref{eq:SeparableInterference} becomes
\begin{eqnarray}
\Psi \Psi^{*}=4 \sin^{2}[\frac{(mK-Mk)\{(x_{1}-x_{2})\}}{(m+M)}].
\label{eq:EntangledInterference}
\end{eqnarray}
This is similar to Gottfried's result for the interference obtained in the correlation between two particles produced in a momentum-conserving decay after each has traversed separate double slits \cite{Gottfried}. Note also that the interference given in eqn. \ref{eq:EntangledInterference} couples particle and mirror variables, illustrating how many-body states interfere with themselves rather than each substate interfering only with itself \cite{silverman}.

Measurement of such a prediction involves preparation of the initial state, adjusting the instruments to measure {\em both} positions $x_{1}$ and $x_{2}$ simultaneously, making the measurement, and then repeating this procedure over an ensemble to build a distribution. 

To illustrate how this leads to a measurement of the mirror fringe spacing, first the apparatus is set to always measure the particle at fixed position $x_{1}$ while the mirrors cm is measured at different positions $x_{2}$ for different members of the ensemble. Using the approximation $m/M<<1$ in eqn. \ref{eq:EntangledInterference} leads to interference fringes \emph{for the mesoscopic/macroscopic mirrors cm} which vary from maximum to minimum through a distance
\begin{eqnarray}
\Delta x_{2} \approx  \pi \hbar/(m(v-V)).
\label{eq:fringespacing}
\end{eqnarray}

Similarly, by always measuring the mirror at fixed $x_{2}$ while varying the measurement position of the particle $x_{1}$ for different members of the ensemble, this approximation leads to interference fringes \emph{for the microscopic particle's cm} which vary from maximum to minimum through a distance $\Delta x_{1}=\Delta x_{2}$. For a static mirror both the mirror and particle fringes are spaced at half the deBroglie wavelength of the {\em particle}, which can be up to $10^{-6}$ m for ultra cold atoms \cite{cronin}.

One issue predicted here, fundamental to extending the quantum-classical boundary, is that of interferometric effects which do not become imperceptible in the limit of large mirror mass. Such effects are surprising in comparison with the imperceptible fringe spacing for a massive particle traversing a double slit.

To understand why fringe spacing in reflection is robust with respect to variations in $M$ consider simplifying the two-body state into two one-body states. The first consists of the wavefunctions of the incident and reflected particle. The second consists of the mirror wavefunctions before and after reflection of the particle, both of which travel in the same direction but with different momenta.

In the case of a mirror whose speed is close to zero, superposition of these particle states results in a ``standing wave'' with nodes or fringes spaced at about half the particle deBroglie wavelength. Superposition of the moving mirror states, on the other hand, forms a fringe spacing commensurate with the difference in momentum between the incident and reflected {\em particle} states. The larger the mass of the mirror the smaller the change in velocity of the mirror upon reflection. Yet the difference in momentum between the mirror before and after reflection remains the same, depending only on the change in momentum of the particle and {\em not} on the mass of the mirror. It is this difference in mirror momenta which leads to the phase difference $\Delta K~x_{2}$ when superposing these two mirror wavefunctions. That is, the mirror fringe spacing is only determined by the change in particle momentum since conservation of momentum in reflection requires $\Delta K=-\Delta k$. While this simplified model, reducing a two-body state into two one-body states, demonstrates why interference for a mesoscopic/macroscopic mirror mass does not become imperceptible as $m/M \rightarrow 0$, it does not account for the correlation given by the exact solution in eqn. \ref{eq:EntangledInterference}, which is inherently a two-body effect.

Double slit interference with a massive particle, on the other hand, superposes two one-body states with the same momentum whose difference in phase, $K \Delta x$, is due to the difference in path lengths from each slit to the measurement point times an extremely large wavevector. This results in an imperceptible fringe spacing.

\subsection{Conservation of probability}

The probability of measuring the particle at $(x_{1},t)$ and the mirror at $(x_{2},t)$ is given by $\iint PDF[x_{1},x_{2},t] dx_{1} dx_{2}$ with the joint PDF determined by the solution of equation \ref{eq:Schreqn} as $\Psi \Psi^{*}$. Using this equation, conservation of probability can then be expressed locally as,
\begin{eqnarray}
\frac{\partial PDF[x_{1},x_{2},t]}{\partial t}+
\frac{\partial j_{1}[x_{1},x_{2},t]}{\partial x_{1}}+\notag \\ \frac{\partial j_{2}[x_{1},x_{2},t]}{\partial x_{2}}=0,
\label{eq:consProb}
\end{eqnarray}
where $j_{1}[x_{1},x_{2},t]=\hbar (\Psi^{*} \partial_{x_{1}} \Psi-\Psi \partial_{x_{1}} \Psi^{*})/(2 i m)$ and  $j_{2}[x_{1},x_{2},t]=\hbar (\Psi^{*} \partial_{x_{2}} \Psi-\Psi \partial_{x_{2}} \Psi^{*})/(2 i M)$. While the expressions for these current densities appear similar to that for one particle systems there are subtle but important differences for a two-body system \cite{currentdensity}.

Multiplying equation \ref{eq:consProb} by $dx_{1} dx_{2}$, integrating over the segment from $a$ to $b$ along the x-axis ($a \leq x_{1} \leq b$ and $a \leq x_{2} \leq b$), and then rearranging terms yields a solution to equation \ref{eq:consProb} if
\begin{eqnarray}
\frac{\partial}{\partial t} \int_{a}^{b} \int_{a}^{b} PDF[x_{1},x_{2},t] dx_{1} dx_{2} +\notag \\ \int_{a}^{b} (j_{1}[b,x_{2},t]-j_{1}[a,x_{2},t]) dx_{2} \notag \\  +  \int_{a}^{b} (j_{2}[x_{1},b,t]-j_{2}[x_{1},a,t]) dx_{1}=0  \notag.
\label{eq:cons2}
\end{eqnarray}
This is most easily interpreted using an $\overline{ab}$ by $\overline{ab}$ rectangular region aligned along the $x_{1}$ and $x_{2}$ axes. The time rate of change in probability within this region is determined by the change in flux of probability out along the $x_{1}$ axis ($j_{1}[b,x_{2},t]-j_{1}[a,x_{2},t]$) plus that along the $x_{2}$ axis $(j_{2}[x_{1},b,t]-j_{2}[x_{1},a,t])$. Since these fluxes in general vary spatially they have to be integrated over this variation along the rectangular boundaries. Numerical integration of the wavegroup results presented below are consistent with conservation of probability.

\subsection{Two-particle wavegroup results: particle-mirror interaction}

\subsubsection{Reflection of wavegroups}
\label{sec:mirror}

To better understand the practical consequences of these results, wavegroups are next formed from a superposition of the incident and reflected `harmonic wavefunctions' (given in eqn. \ref{eq:superposition}) expressed in terms of the correlated particle and mirror substates. An analytic expression for such wavegroups can be obtained for a Gaussian distribution in wavevector components $k$ and $K$. For the mirror, this is proportional to $\exp [-(K-K_{0})^{2}]/(2 \Delta K^{2})$, where the peak of the distribution is at $K_{0}$ and $\Delta K$ is its width, while for the particle, this is proportional to $\exp [-(k-k_{0})^{2}]/(2 \Delta k^{2})$, where the peak of the distribution is at $k_{0}$ and $\Delta k$ is its width. The incident wavegroup propagates in the $(x_{1},x_{2})$ plane along a line whose slope is determined by a ratio of the group velocities of each substate and spreads due to dispersion independently in each direction.

\begin{center}
\begin{figure}
\includegraphics[scale=0.29]{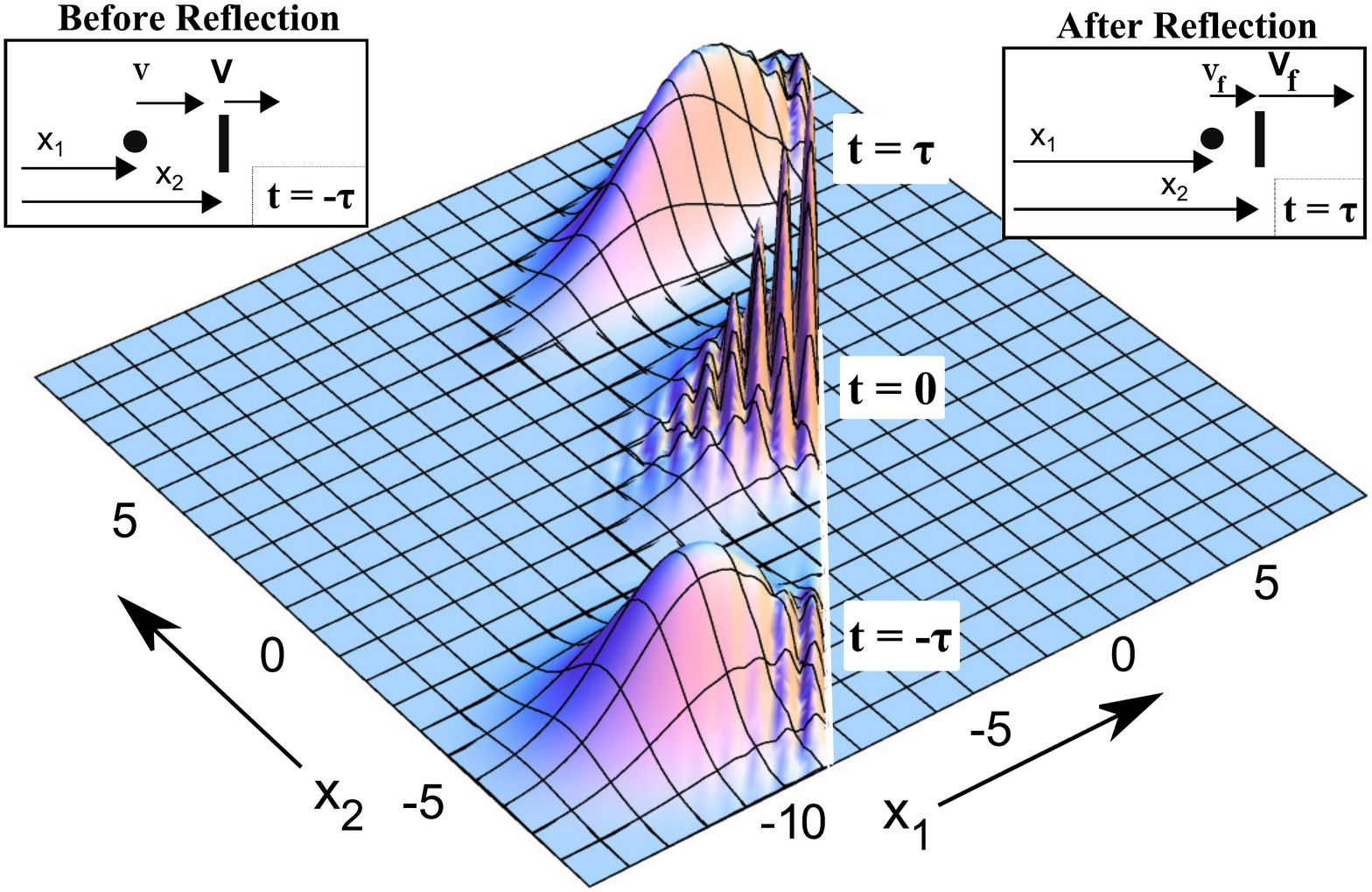}
\caption{Two-body joint probability density snapshots for three sequential times vs coordinates $(x_{2},x_{1})$. The lower PDF waveform moves toward the diagonal white line, corresponding to $x_{1}=x_{2}$, then reflects in the middle snapshot where the incident and reflected two-body wavefunctions `overlap', and finally it moves away from the diagonal in the upper snapshot. The upper left inset is a schematic of the `classical' analog before reflection while the upper right inset is that after reflection with initial and final particle and mirror velocities $v$, $V$, $v_{f}$, and $V_{f}$ respectively. There is no classical analog for the middle snapshot.}
\label{fig:interference}
\end{figure}
\end{center}

\begin{center}
\begin{figure}
\includegraphics[scale=0.3]{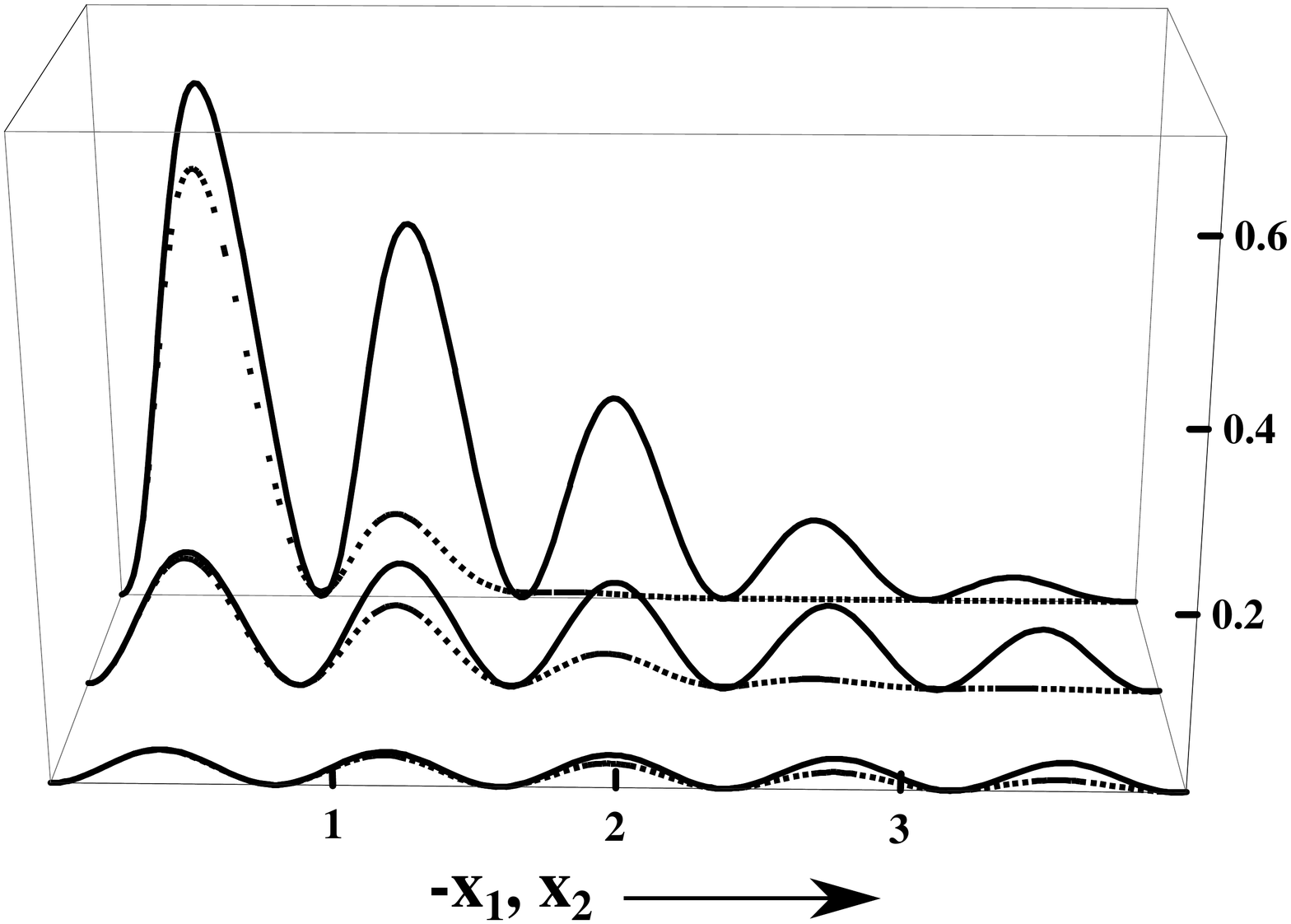}
\caption{Slices of the middle snapshot of fig. \ref{fig:interference} which show the fringe spacing along the $x_{2}$ axis for $x_{1}=0$ (dashed lines) and along the $x_{1}$ axis for $x_{2}=0$ (solid lines). The $x_{1}$ axis has been inverted to display both the dashed and solid lines together. Although each graph has $\Delta K/\Delta k=2$ the value of $\Delta K$ increases sequentially by a factor of $2$ from the front to the back. }
\label{fig:fringespacing}
\end{figure}
\end{center}

In fig. \ref{fig:interference} snapshots of the two-body probability density function are shown at three times for $M/m=100$, $\Delta K/\Delta k=2$, and $K/k=60$. A slice of fig.  \ref{fig:interference} for $x_{1}=0$ along the $x_{2}$ coordinate is shown in fig. \ref{fig:fringespacing} (the solid line) along with a slice of this fig. for $x_{2}=0$ along $x_{1}$ (the dashed line) for different bandwidth wavegroups. This fig. illustrates that the fringe spacing for the particle and mirror substates with narrow bandwidth wavegroups, are essentially the same for the particle and mirror substates, as discussed for the approximation $M/m>>1$ following eqn. \ref{eq:fringespacing}.

Verification of correlated interference requires simultaneous measurement of the particle and cm of the mirror with instruments having a spatial resolution smaller than the fringe spacing. For a static mirror this spacing is half the deBroglie wavelength of the particle, which at $5000$ \AA~for ultracold atoms \cite{cronin} satisfies this requirement while it is dubious at $1.4$ \AA~for slow neutrons \cite{pushin}.

Another constraint on the interference illustrated in fig. \ref{fig:interference} is that the fringe visibility function must be non-zero. That is, the incident and reflected particle-mirror wavegroups must `overlap.' The interference fringes are then determined predominately by a superposition of `harmonic wavefunctions' \cite{Hamilton}. For example, the interference shown in fig. \ref{fig:interference} is determined predominately by eqn. \ref{eq:EntangledInterference} when the wavegroups `overlap' in the center snapshot. The longitudinal coherence lengths for both the particle and mirror are greater than the fringe spacing in this figure. In the upper snapshot there is neither `overlap' nor such interference. The fringe visibility function is non-zero if each wavegroup substate `overlaps' within approximately a longitudinal coherence length \cite{coherencelength}, which is given by $l_{c} \approx \lambda^{2}/\Delta \lambda = \lambda V/\Delta V$ \cite{hasselbach}. For particle substates this can be $l_{c}^{particle}=10000$ \AA~for ultracold atoms \cite{cronin} or $l_{c}^{particle}=790$ \AA~for slow neutrons \cite{pushin}.

If the uncertainty in the mirror velocity is determined by its thermal equilibrium with the environment then $\Delta V_{thermal} \approx \sqrt{2k_{B}T/M}$ yielding $l_{c}^{thermal} \approx h/\sqrt{2Mk_{B}T}$. This expression is consistent with results for ultra-cold atoms in a Bose-Einstein condensate \cite{cronin}.

In the interference region, simultaneous measurement is approximately confined both to a spatial region determined by the two coherence lengths and a temporal region given by the time during which the wavegroups overlap. The former is small for mesoscopic/macroscopic mirror masses while the duration of the interference is essentially determined by the speed of the particle and its coherence length which is approximately equal to $l_{c}^{particle}/v$ for a static mirror when $l_{c}^{particle} >> l_{c}^{mirror}$. One method, discussed below, to reduce these coherence length limitations is with multiple such two-body states. First however, the effect of variations in mirror coherence lengths on correlated interference is illustrated.

\subsubsection{Mirror coherence length variation}
\label{sec:bandwidth}

\begin{center}
\begin{figure}
\includegraphics[scale=0.26]{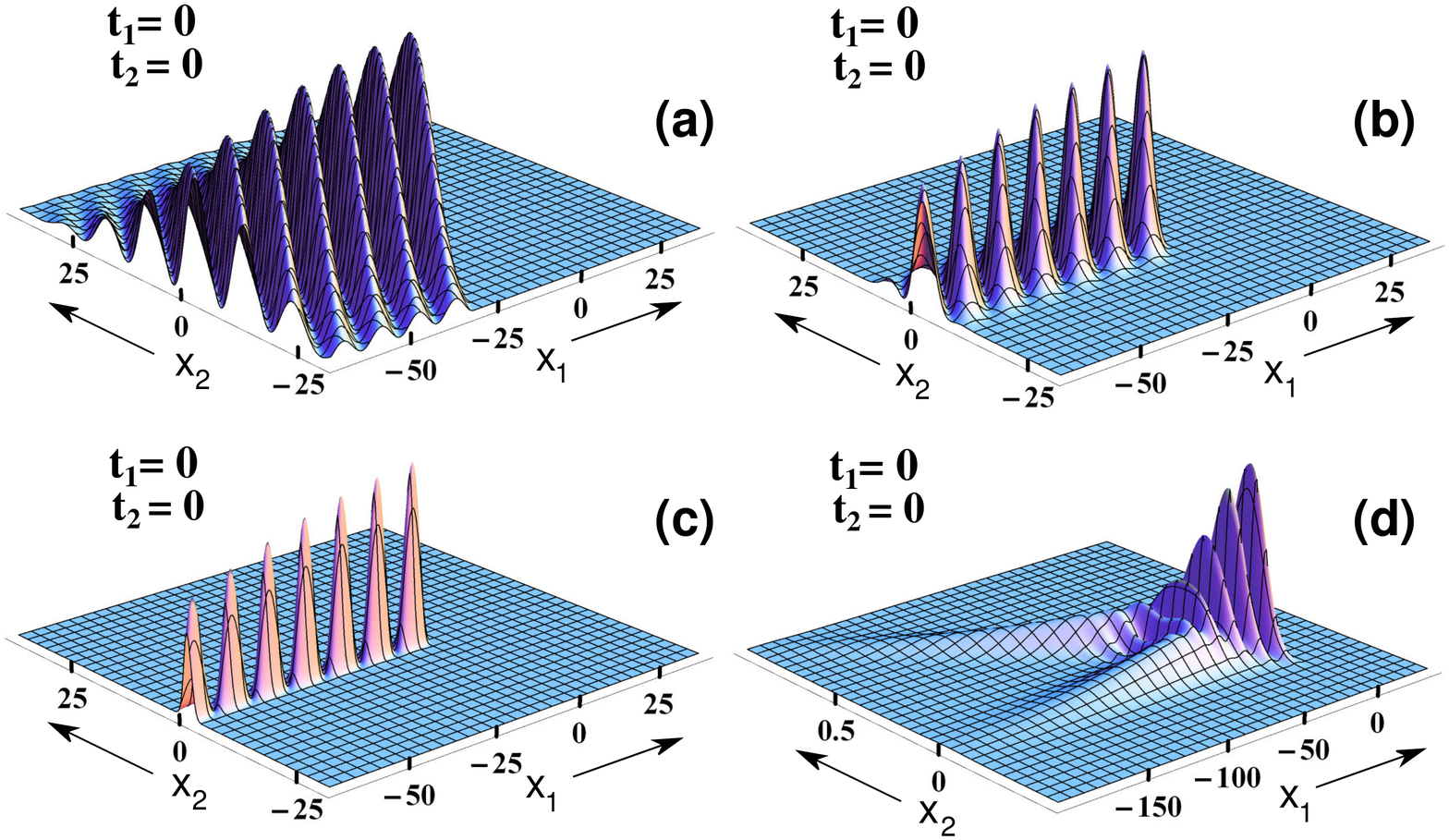}
\caption{Two-body PDF plots similar to the center snapshot of fig. \ref{fig:interference} but with different spreads in mirror velocities for a fixed spread in particle velocities while all other parameters are the same. $\Delta V/\Delta v=80, 20, 5$, and $0.4$ for $M/m=200$ in parts a, b, c, and d respectively. Note the change of scale in graph d.}
\label{fig:spreadvelocity}
\end{figure}
\end{center}

Fig. \ref{fig:spreadvelocity} illustrates how variation in the longitudinal coherence length of the mirror substate affects the particle-mirror interference with fixed particle substate coherence length. Part a shows a longer mirror coherence length than is used in fig. \ref{fig:interference} while parts \ref{fig:spreadvelocity}b through \ref{fig:spreadvelocity}d progressively reduce the coherence length of only the mirror substate. One might expect that the small coherence length associated with a mesoscopic/macroscopic mirror mass would not allow for the interference shown in fig. \ref{fig:spreadvelocity}a or \ref{fig:spreadvelocity}b.

In fig. \ref{fig:spreadvelocity} d a small mirror coherence length along with a large mirror recoil prevents wavefunction overlap over a range of $x_{1}$ values ($x_{1}<-100$) where it was present before. Nevertheless, a slice along the $x_{2}$ axis for measurement of the particle at these values of $x_{1}$ indicates a splitting of the mirror substate into two states which do not overlap and are therefore distinguishable. 

This splitting is a consequence of two ways that the particle could have reached $x_{1}$. It could have come from the incident {\em or} reflected particle wavegroup substates since the position of the particle is not known to within its large coherence length. As the {\em mirrors} coherence length increases, the wavefunctions associated with these two ways overlap and generate correlated interference as shown in fig. \ref{fig:spreadvelocity}a. As the mirrors coherence length decreases, the position of the mirror before reflection is distinguishable from that after reflection due to mirror recoil, which results in {\em no interference of either the particle or mirror}, as shown in fig. \ref{fig:spreadvelocity}d for $x_{1}<-100$. That is, measurement of the particle at  $x_{1}=-100$ and mirror at  $x_{2}=0$ is consistent with conservation of momentum for the particle and mirror not to have reflected while measurement of the particle at  $x_{1}=-100$ and mirror at  $x_{2}=0.5$ is consistent with conservation of momentum for the particle and mirror to have reflected. The short coherence length of the mirror relative to the recoil distance then results in distinguishable two-body states. The condition for interference in a two-body system therefore involves the coherence of both bodies. 

\subsubsection{Coherence transfer: particle-mirror interaction}
\label{sec:coherencetransfer}

\begin{center}
\begin{figure}
\includegraphics[scale=0.28]{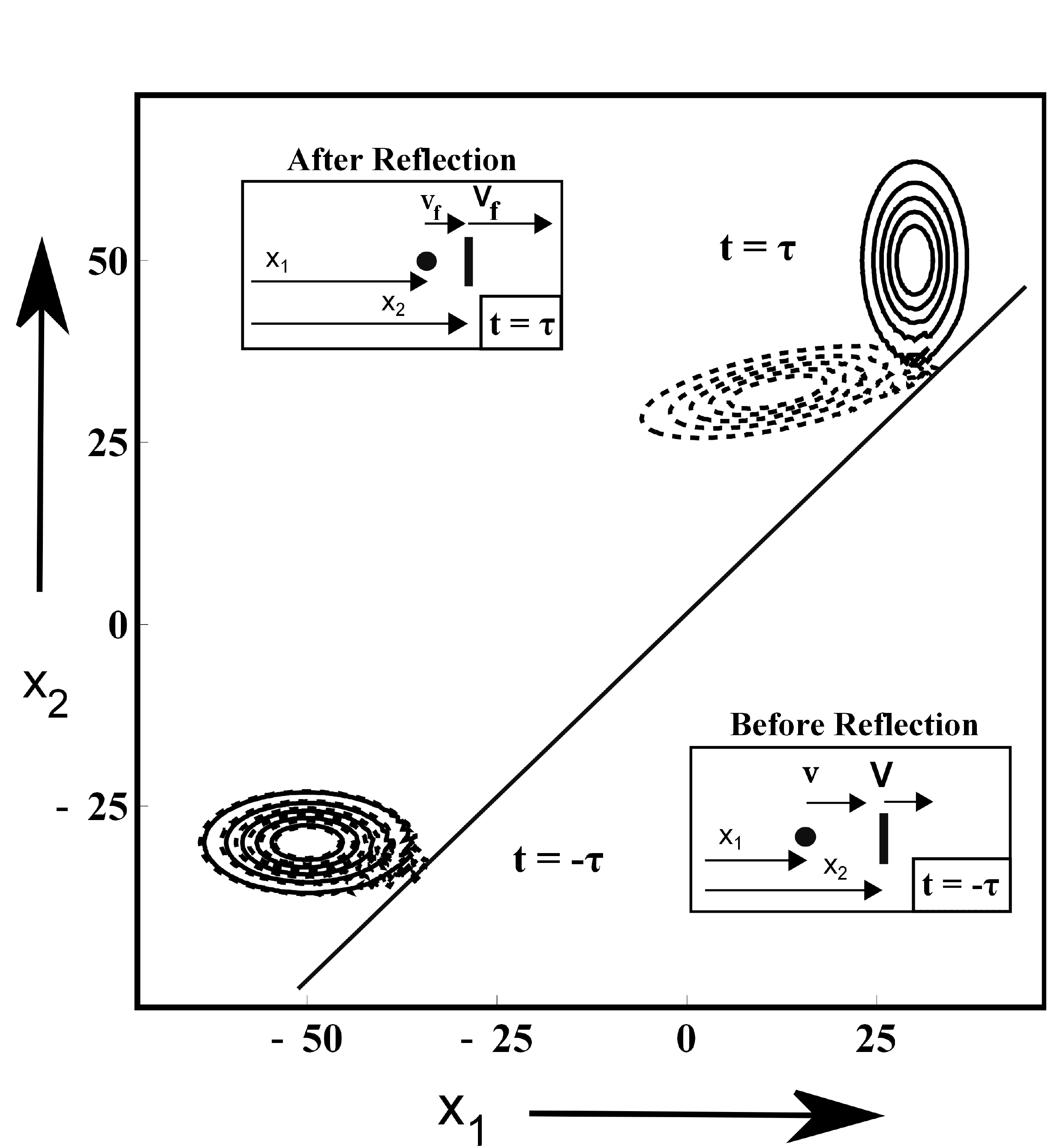}
\caption{Two-body contour plots of the joint probability density snapshots similar to fig. \ref{fig:interference} but without the interference or ``overlap'' region snapshot, illustrating coherence transfer for two different particle-mirror mass ratios. The solid and dashed lines are for a spread in velocities given by $\Delta V/\Delta v=10$ with $M/m=1$ and $M/m=20$, respectively.}
\label{fig:coherencetransfer}
\end{figure}
\end{center}

After reflection the spatial width of the mirror wavegroup substate is exchanged with that of the particle wavegroup substate when $M=m$. This is most easily seen by constructing a particle-mirror wavefunction with different bandwidths for the particle and mirror wavegroup substates, shown in fig. \ref{fig:coherencetransfer} which is a contour plot of joint PDFs similar to figure \ref{fig:interference} but without the middle snapshot in the interference region. The solid and dashed contours correspond to $M/m=1$ and $M/m=20$ respectively with the spread in velocities given by $\Delta V/\Delta v=10$.

This result can be understood by comparing classical and quantum reflection. In a one-dimensional classical collision, conservation of energy and momentum require an exchange of particle-mirror velocities independent of either velocity for $m=M$. This is manifest quantum mechanically in the exchange of commensurate  wavefunction parameters $k$ and $K$ between the incident and reflected two-body wavefunctions. 

If an incident particle substate, consisting of only one harmonic component (corresponding to speed $v$) reflects from a mirror substate with many velocity components, then each harmonic component of the mirror substate (corresponding to different values of $V$) reflects the particle substate and therefore acquires velocity $v$ while the reflected particle substate acquires different velocity values for each reflected component of the mirror substate. This results in the reduction of the mirror bandwidth and an increase in the particle bandwidth, which is manifest in fig. \ref{fig:coherencetransfer} as the exchange of incident and reflected wavegroup shapes. It also is responsible for the distortion of the reflected wavegroup shape in fig. \ref{fig:interference}.

Experimentally confirming coherence transfer involves reflection with disparate coherence lengths. On a microscopic scale, this could be observed between an atom which has been cooled to ultracold temperatures retro-reflecting from the same type of atom which has a shorter coherence length. The transfer of coherence might then be monitored not by a measurement of the PDF but rather by its coherence properties via interferometry. A more useful implementation for increasing the coherence length of larger objects could involve a group of Bose-Einstein particles (the condensate), whose total mass is similar to that of the mirror, retro-reflecting`coherently' from a mirror as `one particle' rather than a collection of individual particles.

\section{Multiple two-body states: overview}
\label{sec:twobodyoverview}

To mitigate both spatial and temporal fringe localization in reflection due to small particle and mirror coherence lengths, consider interference of multiple such two-body states: a one-dimensional example of which is the interference from the reflection of a neutron by each of the two surfaces of a moving aluminum `slab'. The two-body PDFs associated with two models of such reflection are calculated below. The first, presented in the next section, simply approximates each weak surface reflection as being due to a delta function potential which allows for both transmission and weak reflection. These two delta function potentials are fixed to and symmetrically offset from the cm of the slab (of length $D$) by a distance $D/2$, with multiple reflections neglected. 

The results from this simple model then corroborate those from a more accurate model, given in section \ref{sec:slab}, involving a two-body treatment of a particle interacting with a finite barrier or finite well. The delta function `beamsplitter' model is included to focus attention on the physics associated with generating multiple two-body states. The more accurate finite barrier or well treatments conceal the fundamental physics but yield similar results. In essence, the latter incorporates multiple reflections along with the appropriately modified wavevectors between the surfaces of the slab while the former does not. The results of the simpler delta function model approach those of the finite barrier-well model as the reflections from the slab interfaces become weaker. 

For completeness, the results of the particle trapped in a moving infinite well are included in subsection \ref{sec:well}. While the finite barrier or well examples discussed below deal only with unbound states, the infinite well is an example of the two-body formalism for a bound state.

The classical analogy for the neutron reflecting from each of the two surfaces of a moving aluminum `slab' is that of a short pulse of light reflecting from a moving thin film. As the film's thickness, speed, or the average wavelength of the incident pulse changes, the reflected pulse intensity then varies as a consequence of the interference between the reflected waves from the two interfaces. 

Classically, such thin film interference mitigates both the spatial and temporal fringe localization associated with reflection of a pulse of light from a moving mirror (the analog of the two-body quantum treatment of a particle reflecting from a moving mirror). Interference between the incident and reflected two-body pulse from the moving mirror is localized spatially by the coherence length of the pulse. These fringes are also temporally constrained by the pulse duration. 

However, interference of the two reflected light pulses, one from each surface of the thin film, dramatically reduces this fringe localization for two reasons: first the two reflected light pulses travel in the same direction thereby maintaining spatial overlap and second they experience the same Doppler shift in reflecting from the two interfaces which move at the same speed resulting in fringes which are not time dependent (this is not the case for interference of the incident and reflected pulses from a moving mirror). 

For similar reasons, interference of the two-body wavegroup states associated with reflection from each surface of the slab mitigates the fringe localization associated with reflection from a mirror. However, the two-body interference is now between two wavegroups, each analogous to the one shown in the upper snapshot of fig. \ref{fig:interference} and each of which corresponds to a reflection from only one surface of the slab and both of which move in the same direction thereby maintaining overlap.

The classical treatment yields interference only for the pulse of light and not the mirror from which it reflects. The two-body quantum treatment presented below yields predictions about correlations in the measurements of both the neutron and slab. Destructive interference of these wavegroups then corresponds to no probability of measuring the neutron and slab along the overlap region. A change in the slab's thickness, speed, or the average wavelength of the incident particle then results in a variation of the probability of simultaneously measuring the neutron and slab between zero and some maximum value, corresponding to constructive and destructive interference.

A pulse of light reflecting from a thin film can also generate a standing wave within the film under appropriate conditions. Similarly, two-body wavefunctions also generate such a ``Fabry-P\`erot'' resonance which is illustrated below.

\section{Multiple two-body states: Modeling the slab surfaces with two delta function potentials}
\label{sec:twodelta}

\subsection{Method}
\label{subsec:method}

Modifying $\beta$ to split rather than totally reflect the incident wave while including an offset in the delta function argument to model the surface which is offset from the cm of the slab yields the two-body reflected wavefunction. This eigenstate of energy or `harmonic solution', for each surface reflection, is then superposed to form wavegroups associated with refection from that surface. Rather than the waveform of the center snapshot in fig. \ref{fig:interference}, a waveform analogous to that in the upper snapshot is the focus of this discussion. There are now two such reflected neutron-slab wavegroups, one for each slab surface, which interfere.

Since the neutron either reflects from the first surface or transmits through the first to the second, from which it then reflects, both reflected neutron substates have the same distribution of velocities. Since in each case momentum is transferred to the slab only once for each of these two possible paths (reflection from either the first {\em or} second surface of the slab), two associated slab substates are generated, each with the same velocity distribution. This results in two reflected neutron-slab wavegroups traveling at the same speed and direction but offset. 

Interference then requires wavegroup `overlap' within a coherence length of both the neutron and slab substates. An example of a lack of overlap in the mirror substate, resulting in no two-body interference, was discussed in section \ref{sec:bandwidth} and illustrated in fig. \ref{fig:spreadvelocity}d.

This requirement, applied to the slab substate, is that the cm of the two reflected slab substates separate no further than their coherence lengths. This offset is estimated by the extra slab speed, beyond $V$, acquired from the first reflection multiplied by the difference in particle reflection times between the two slab surfaces. Assuming $m/M<<1$ and $V/v<<1$ this offset, $2mD/M$, then must be less than $l_{c}^{slab}$, which if determined by thermal equilibrium requires 
\begin{equation}
T<h^{2}M/(8D^{2}k_{B}m^{2}).
\label{eq:SlabTemp}
\end{equation}

Next consider the constraint on $l_{c}^{neutron}$. When a neutron wavegroup of spatial width $w$ (where dispersion in reflection is neglected) retro-reflects a distance $L$ from the cm of the slab, with $L$ much greater than both the wavegroup width $w$ and the slab length $D$ ($L>>w>>D$), `overlap' of the neutron wavegroups which have reflected from both surfaces occurs. For a near stationary slab the reflected neutron wavegroups are offset by about a distance $2D$, interference from which requires a neutron coherence length exceeding $2D$. An example which satisfies this constraint is a neutron of coherence length $l_{c}^{neutron} \approx 8\times 10^{-8}$~m \cite{pushin} reflecting from the two surfaces of an aluminum slab of thickness $D= 10^{-8}$~m (there is little constraint on the area of this slab and therefore its mass).

These parameters easily satisfy the overlap constraints for both the mesoscopic/macroscopic slab at room temperatures and the neutron. Having achieved overlap, the resulting interference is then determined essentially by the harmonic wavefunctions yielding the {\em constant} two-body reflected joint probability density
\begin{equation}
P \approx \sin^{2}[2DmM(V-v)/(\hbar (m+M))].
\label{eq:SlabInterference}
\end{equation}
Assuming $m/M<<1$ and $V/v<<1$ yields
\begin{equation}
P \approx \sin^{2}[2Dmv/\hbar].
\label{eq:SlabIntApprox}
\end{equation}

The slab {\em and} reflected neutron will never be simultaneously observed when $P=0$ \cite{liu}. Such non-local two-body interference is similar to that of a pulse of light retro-reflecting from a thin film where destructive interference depends neither on the locations of the detector nor the thin film. Indeed, in the limits just given these two cases are mathematically identical if the photon wavelength is replaced with the particle's deBroglie wavelength.

\subsection{Delta function model predictions}
\label{subsec:results}

\begin{center}
\begin{figure}
\includegraphics[scale=0.35]{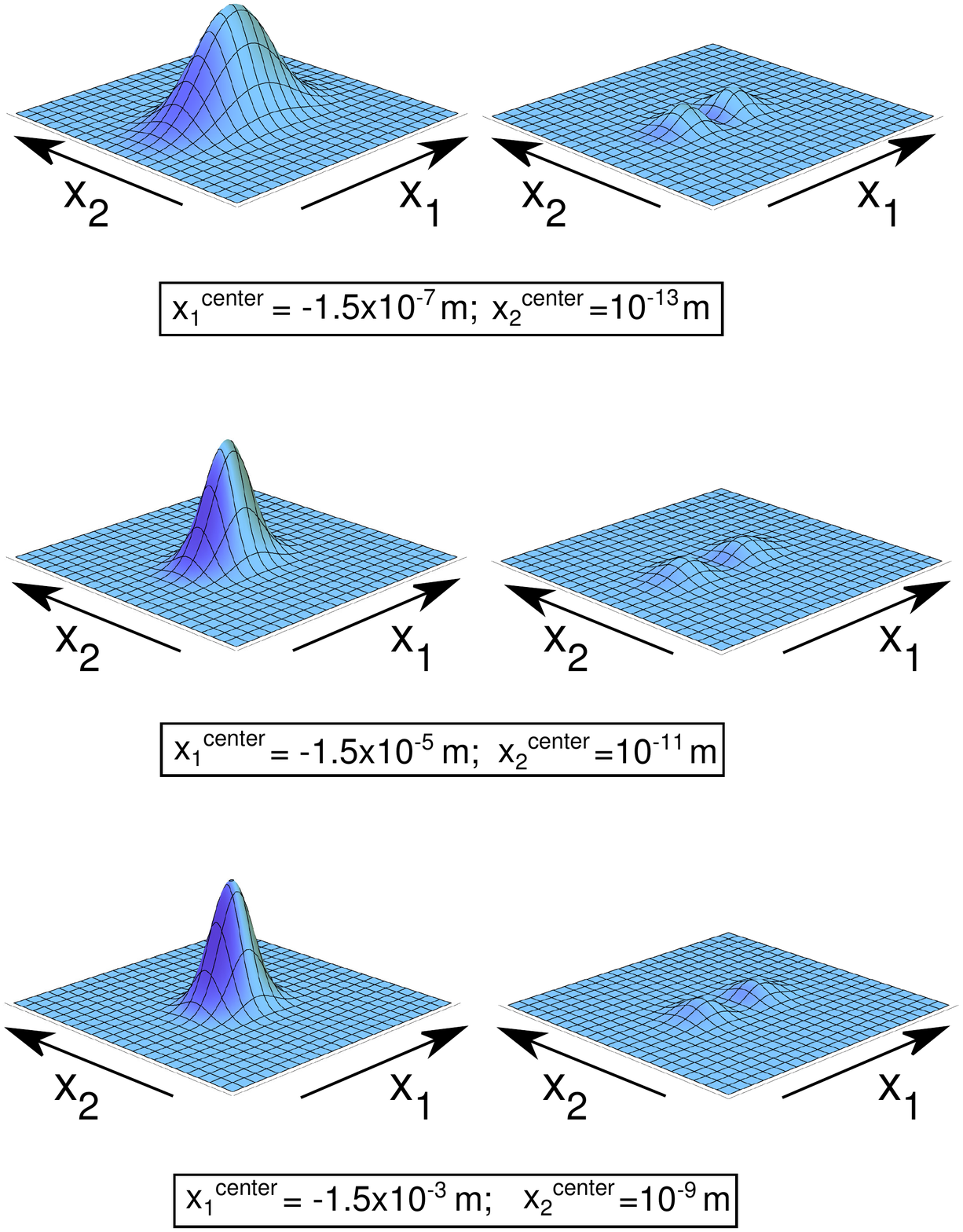}
\caption{Joint probability density snapshots for the neutron-slab at three times vs slab and neutron coordinates ($x_{2}$ and $x_{1}$ respectively) after reflection. The left column graphs differ from the right only in the speed of the incident neutron, illustrating constructive and destructive interference respectively.}
\label{fig:slab}
\end{figure}
\end{center}

The PDF for reflection from the slab is illustrated for wavegroups in fig. \ref{fig:slab} using parameters similar to those above: $l_{c}^{neutron} \approx 8\times 10^{-8}$m, $D=10^{-8}$m, $V=0.001$m/s, $M=10^{-13}$kg, and a slab coherence length determined by $l_{c}^{thermal}$ at $T=1 K$. The neutron reflects from the first slab surface at $x_{2}=0$, $x_{1}=0$, and $t=0$. To illustrate the non-local nature of the interference, three snapshots are chosen for times which allow the center of the wavegroup to change position by four orders of magnitude in both the neutron and slab coordinates.  Since the spatial size of the neutron-slab PDF is limited by small coherence lengths, a graph with a spatial region encompassing all of these points, while still showing the wavegroup structure, cannot easily be generated. Instead `blow-ups' in the $(x_{1},x_{2})$ plane of small regions around these points are shown. The snapshots in the upper row have a span in $x_{1}$ of $1.6 \times 10^{-7}$m and a span in $x_{2}$ of $2 \times 10^{-14}$m with the approximate centers of each graph located at $x_{1}=-1.5 \times 10^{-7}$m and $x_{2}= 10^{-13}$m. The middle row of graphs have a span in $x_{1}$ of $2.9 \times 10^{-7}$m and a span in $x_{2}$ of $2 \times 10^{-12}$m with the approximate centers of each graph located at $x_{1}=-1.5 \times 10^{-5}$m and $x_{2}= 10^{-11}$m. The lower row of graphs have a span in $x_{1}$ of $2.9 \times 10^{-5}$m and a span in $x_{2}$ of $2 \times 10^{-10}$m with the approximate centers of each graph located at $x_{1}=-1.5 \times 10^{-3}$m and $x_{2}= 10^{-9}$m. The increase in span is needed since the wavegroup expands over these time scales. The vertical scale therefore decreases from top to bottom rows but is the same within each row so that PDF heights can only be compared within a row.

The only parameter that differs between the graphs in the left and right columns of fig. \ref{fig:slab} is the neutron velocity before reflection. It is $v=1448$ m/s for the left and $v=1458$ m/s for the right column respectively. These two neutron speeds result in either constructive or destructive interference (left or right columns respectively) of the neutron-slab wavegroups which are offset by $l_{c}^{neutron}/5$ due to reflection from the two slab surfaces. As this offset decreases (e.g. by decreasing $D$) or $l_{c}^{neutron}$ increases, the PDF goes to zero approaching the destructive interference predicted in eqn. \ref{eq:SlabIntApprox} (since $m/M<<1$ and $V/v<<1$). Similar calculations also confirm the predictions that there is little effect on the interference from increasing $M$ while changes in $v$ have a significant effect. Note also that the small size of the slab coherence length, $l_{c}^{slab}=4 \times 10^{-16}$m, is much larger than the offset due to reflection from the two slab surfaces, $3 \times 10^{-22}$m, and therefore has little effect on the destructive interference.

Increasing the mass of the slab changes only the span in the $x_{2}$ coordinate of these graphs but not the interference. The parameters in fig. \ref{fig:slab} are {\em not} chosen to illustrate effective wavegroup destructive interference but rather to show coherence length effects on the interference using experimentally realizable values.

\section{Multiple two-body states:  Modeling reflection from a slab with a well or barrier}
\label{sec:slab}

\subsection{Overview}
\label{sec:wellbarriertheory}

The model for a neutron traversing a slab of matter which incorporates multiple reflections and variation in the wavevector inside the slab is determined from the Schr\"odinger equation given by
\begin{eqnarray}
(\hbar^{2} \partial_{x_{1}}^{2}/2m+\hbar^{2} \partial_{x_{2}}^{2}/2M+PE[x_{1}-x_{2}] \notag \\
+i\partial_{t})\Psi[x_{1},x_{2},t]=0,
\label{eq:Scheqn}
\end{eqnarray}
where $PE[x_{1}-x_{2}]$ is now the potential energy associated with either the well or barrier. Of interest is a solution which yields an energy eigenstate for the particle-well or barrier interaction, for which neither the particle nor the well or barrier is localized. Wavegroups are then constructed from these states. 

A separable solution to this two-body Schr\"odinger equation is obtained in a manner similar to that outlined in section \ref{sec:theoryparticlemirror}. The center of mass and relative coordinate transformation yields solutions which are matched at the boundaries and then transformed to the particle-well coordinates. The infinite well potential is presented to illustrate the application of the two-body formalism to a bound state. Then in subsection \ref{subsec:barrier} calculations are outlined for a particle interacting both with a finite barrier and a finite well in unbound states. These provide a more accurate model for a neutron traversing a slab of matter than that just discussed.

\subsection{Infinite well energy eigenstates: correlated two-body bound states}
\label{sec:well}

The boundary condition for the infinite potential well, of width $D$, is $\Psi[x_{rel} \pm D]\rightarrow 0$. The relative wavefunction does not exist outside the well for $x_{rel}<-D$ and $x_{rel}>D$. A solution to eqn. \ref{eq:Scheqn} for a particle which has reduced mass is then \cite{serge}
\begin{equation}
\Psi_{rel}  \propto e ^{-i E_{rel} t/\hbar } \sin[\frac{n \pi (x_{rel}+D)}{2D}]~\Theta[x_{rel},D],
\label{eq:Psiwell2}
\end{equation}
where $\Theta[x_{rel},D]=(\theta [x_{rel}+D]-\theta [x_{rel}-D])$ has value one within the well and zero everywhere else with n being the number of nodes.

The solution to the center of mass ODE (the same as eqn. \ref{eq:ScheqODE1}) is 
\begin{equation}
\Psi_{cm} \propto e ^{i (K_{cm} x_{cm}-E_{cm} t/\hbar }).
\label{eq:Psiwell1}
\end{equation}
The complete solution is then $\Psi[x_{cm},x_{rel},t] \propto \psi_{cm}\psi_{rel}$.

Again, we are interested in measurements of the particle and well rather than measurements of two transformed ``objects,'' one with a reduced mass and the other with the total mass of the system. Expanding the sine function in eqn. \ref{eq:Psiwell2} into exponential form results in wavefunctions traveling in opposite directions. Transforming back to the particle-well coordinates, the momenta and energies of the particle and well differ in magnitude in these two directions. 

The modes of the well, characterized by the number of nodes $n$, are determined in the cm-rel coordinates by $K_{rel}=n\pi/2D$. In the particle-well coordinates this constrains the incident values of particle and well velocities. That is, given an initial well velocity $V$ only the following initial particle velocities are allowed,
\begin{equation}
v=V+n \pi \frac{\hbar (m+ M)}{2DmM}.
\label{eq:quantizationvV}
\end{equation}

In the particle-well coordinates the wavefunction's phase, $\phi$, needs to have its temporal part parsed into the kinetic energy of the particle and the kinetic energy of the well. This is done as in the previous calculations. The resulting kinetic energies of the particle and well differ as the particle moves in opposite directions. These energies and their associated momenta are those expected of a classical particle reflecting from the moving well walls.

\subsection{Finite barrier-well energy eigenstates: correlated two-body unbound states}
\label{subsec:barrier}

The finite barrier or well (referred to as only the barrier) has potential energy $PE$ (positive for the barrier and negative for the finite well) and extends over a distance $D$. This divides space into three regions: before the barrier or ``before'', in the barrier region or ``barrier,'' and after the barrier or ``after.'' Solutions are first obtained for these three regions in the cm and rel coordinates by solving eqns. \ref{eq:ScheqODE1} and \ref{eq:ScheqODE2}.

The solution to eqn. \ref{eq:ScheqODE2} before the barrier consists of incident and reflected wavefunctions given by
\begin{eqnarray}
\Psi_{rel}^{before} = A e^{i (K_{before} x_{rel}-E_{rel} t/\hbar }) \notag \\ +B e^{i (-K_{before} x_{rel}-E_{rel} t/\hbar }),
\label{eq:Psibefore}
\end{eqnarray}
where $K_{before}=\sqrt{2\mu E_{rel}}/\hbar$. The solution in the barrier-finite well region also consists of incident and reflected wavefunctions given by
\begin{eqnarray}
\Psi_{rel}^{barrier} = F e^{i (K_{barrier} x_{rel}-E_{rel} t/\hbar }) \notag \\ +G e^{i (-K_{barrier} x_{rel}-E_{rel} t/\hbar }),
\label{eq:Psibarrier}
\end{eqnarray}
where $K_{barrier}=\sqrt{2\mu (E_{rel}-PE)}/\hbar$. The solution after the barrier consists only of a transmitted wavefunction given by
\begin{eqnarray}
\Psi_{rel}^{after} = H e^{i (K_{after} x_{rel}-E_{rel} t/\hbar }),
\label{eq:Psiafter}
\end{eqnarray}
where $K_{after}=\sqrt{2\mu E_{rel}}/\hbar$.

The boundary conditions are continuity of the wavefunctions and their derivatives with respect to $x_{rel}$ at $x_{rel}=\pm D$. These then constrain the coefficients $B$, $F$, $G$, and $H$ with $A=1$. 

Again, the solution to eqn. \ref{eq:ScheqODE1} is given by eqn. \ref{eq:Psiwell1}. The complete solution is then $\Psi[x_{cm},x_{rel},t] \propto \psi_{cm}\psi_{rel}$. Since we are interested in predictions about measurements of the particle and barrier rather than measurements of the reduced mass and total mass ``objects,'' a transformation from the relative and center of mass to the particle-barrier coordinates is required.

In the particle-barrier coordinates the wavefunction's phase, $\phi$, needs to have its temporal part parsed into the kinetic energy of the particle and the kinetic energy of the barrier. This procedure is the same as it is for the particle-mirror and particle-infinite well two-body states as described above. However, the product of the potential energy part of the total energy with time $t$ in the barrier region must be added to the phase. Such a term has no measurable effect as will be shown next. The wavefunction in the barrier, expressed in the particle-well coordinates, is then 
\begin{eqnarray}
\Psi^{barrier} = F e^{i \{\Phi_{spatial}^{right}-(KE_{1}^{right} t+KE_{2}^{right} t+PE~t)/\hbar\}} \notag \\
 +G e^{i \{\Phi_{spatial}^{left}-(KE_{1}^{left} t+KE_{2}^{left} t+PE~t)/\hbar\}},\notag
\label{eq:Psibarrierlab}
\end{eqnarray}
where $\Phi_{spatial}^{right}$ and $\Phi_{spatial}^{left}$ contain the spatial terms in the phase and are functions of $m,M,v,V,PE,x_{1},x_{2},$ and $\hbar$. The temporal terms contain the kinetic energy for the particle and barrier moving to the right and left, $KE_{1}^{right},KE_{2}^{right},KE_{1}^{left},KE_{2}^{left}$ and the potential energy $PE$. The kinetic energy terms are functions of $m,M,v,V,PE$, and $\hbar$. 

The potential energy term $e^{i PE~t/\hbar}$ is a common factor of both the incident and reflected wavefunctions in the barrier. Since the PDF is generated from the wavefunction multiplied by its complex conjugate, such common factors have no effect on the PDF. There is then no need to associate the potential energy part of the total energy with either the particle or the barrier or as a separate term in the phase, $PE~t/\hbar$. The potential energy part of the total energy has observable consequences only parametrically within the momenta and kinetic energies of the particle and barrier. This simple division of the momentum and energy of the particle in the barrier can be contrasted with that of the stress-energy tensor for an electromagnetic wave traversing a dielectric slab \cite{loudon}.

\subsection{Wavegroup results: overview}
\label{sec:wavegroupoverview}

Wavegroups are next formed from a Gaussian superposition of the two-body energy eigenstates for the particle and infinite well or finite particle and barrier-well described above. Correlated interference is a consequence of any such superposition. However, we focus the following discussion on the subsets of correlated interference effects which deal with the superposition of two such wavegroups. In particular, emphasis is given to the two-body quantum analogy of a pulse of light reflecting from a thin film: interference from the two-body wavegroups which have reflected from the two surfaces of the slab.  At the risk of distracting from this main topic, correlated interference from multiple reflections when such a wavegroup interacts with an infinite barrier are also discussed.

\subsubsection{Particle finite barrier-well wavegroups}
\label{sec:wavegroupbarrierwell}

The wavegroup for the particle and barrier or finite well is calculated using a Gaussian superposition of the energy eigenstates given in subsection \ref{subsec:barrier}. Unfortunately, the coefficients $B$, $F$, $G$, and $H$ of eqns. \ref{eq:Psibefore}, \ref{eq:Psibarrier}, and \ref{eq:Psiafter} depend in a non-trivial manner on the variables of integration. The resulting integrals cannot be determined in closed form. To facilitate the calculation, the following sums will replace these integrals:
\begin{eqnarray}
\Psi_{barrier}^{wavegroup} \propto  \sum_{V_{i}}^{V_{f}} \frac{e^{\frac{-(V-V_{0})^{2}}{2\Delta V^{2}}}}{\sqrt{\Delta V}} \Psi[x_{1},x_{2},t] \notag
\label{eq:barrierwavegroup}
\end{eqnarray}
where the peak of the barrier velocity distribution is at $V_{0}$, $\Delta V$ is its width, and the sum is from an initial barrier velocity $V_{i}$ to a final velocity $V_{f}$. Summing over the particle velocity distribution yields the wavefunction for the wavegroup given by
\begin{eqnarray}
\Psi_{total}^{wavegroup} \propto  \sum_{v_{i}}^{v_{f}} \frac{e^{\frac{-(v-v_{0})^{2}}{2\Delta v^{2}}}}{\sqrt{\Delta v}} \Psi_{barrier}^{wavegroup}, \notag
\label{eq:totbarrierwavegroup}
\end{eqnarray}
where the peak of the particle velocity distribution is at $v_{0}$, $\Delta v$ is its width, and the sum is from an initial particle velocity $v_{i}$ to a final velocity $v_{f}$. Two cases are now treated: the unbound two-body wavegroup state of a particle and finite well and a similar unbound state of a particle and finite barrier.

\paragraph{Particle traversing a finite well: $KE_{initial}^{relative}>PE$.}
\label{sec:barriergreater}

\begin{center}
\begin{figure}
\includegraphics[scale=0.29]{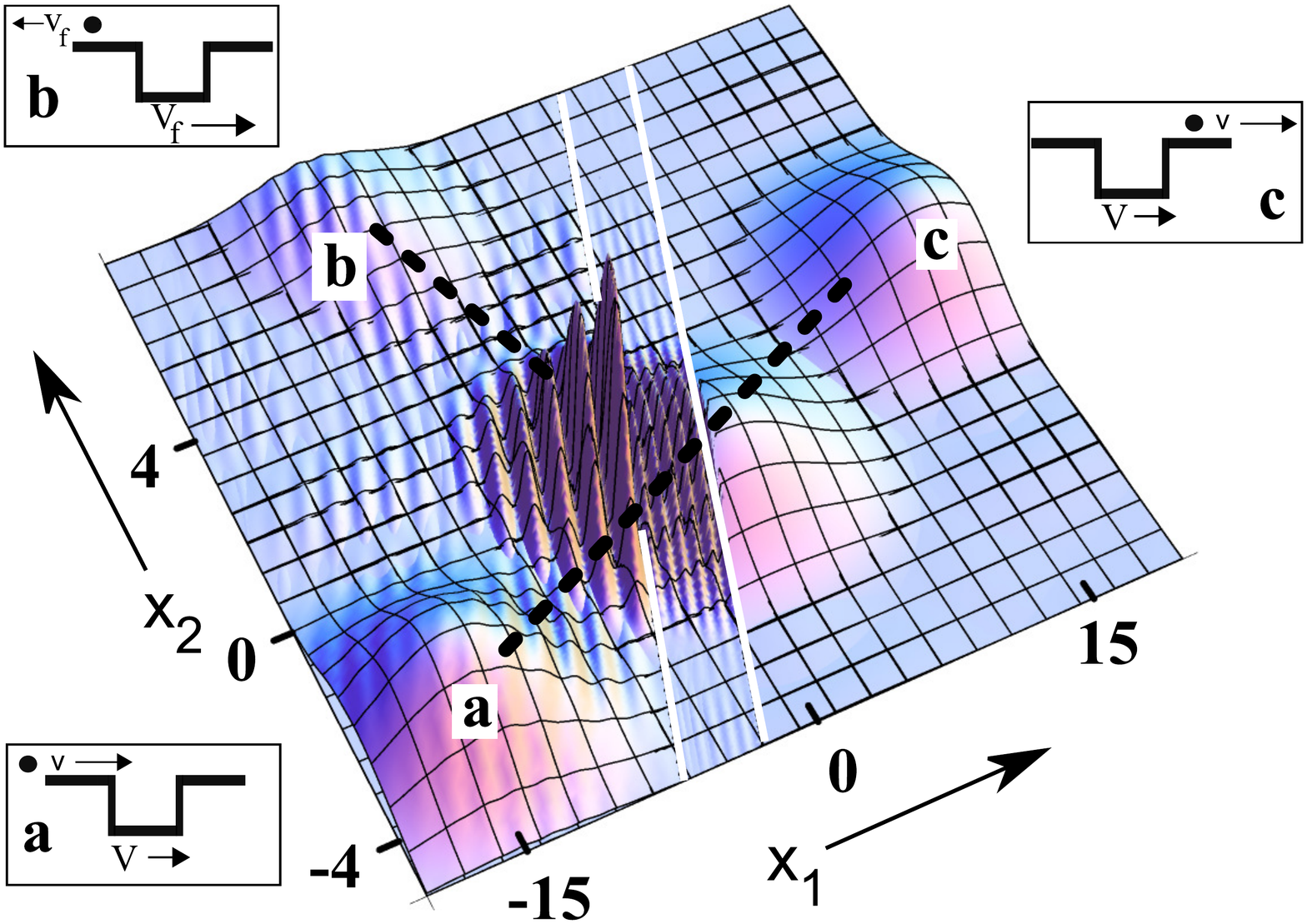}
\caption{PDF snapshots for three sequential times vs coordinates $(x_{2},x_{1})$. The first snapshot generates peak $a$ while peaks $b$ and $c$ comprise the last snapshot. The classical analogs of these peaks are shown in the insets. The PDF progresses temporally along the dashed line.}
\label{fig:barrier_negenergy}
\end{figure}
\end{center}

Consider next a particle interacting with a finite well whose sum of initial kinetic energies in the relative coordinate system is greater than the well PE. The size of the particle substate wavegroup is chosen to be a few times larger than the finite well width $D$. Fig. \ref{fig:barrier_negenergy} shows results of the PDFs for three sequential snapshots taken at equal time intervals progressing along the dashed line from the lower left to upper right and upper left. The analogous classical positions of the particle and finite well for particular snapshots are illustrated in the insets. The barrier boundaries occur at the diagonal white lines, corresponding to $x_{1}=x_{2} \pm D$. The parameters used in fig. \ref{fig:barrier_negenergy} are $v_{0}/V_{0}=6$, $\Delta v/\Delta V=1.5$, and $M/m=5$ while the $KE_{initial}^{relative}-PE/\mid PE\mid =1.4$ using the average value of $KE_{initial}^{relative}$ for the wavegroup particle and well distributions. 

One category of correlated interference, which we call type I, occurs when the incident and reflected two-body wavefunctions, traveling in opposite directions in the $(x_{2},x_{1})$ plane, `overlap.' This is illustrated in fig. \ref{fig:barrier_negenergy} for weak reflection by the fringes of the middle snapshot just to the left of the barrier or line $x_{2}=x_{1}-D$ and also by the fringes in fig. \ref{fig:interference}. These fringes are spaced by about half the deBroglie wavelength of the {\em particle} for $M>>m$ and $v>>V$ as are the similar correlated interference fringes in the two-body reflection of a particle from a mirror given in eqn. \ref{eq:EntangledInterference}. 

However, the interaction generates another form of correlated interference when the reflected wavegroups, one from each barrier surface, interfere as they travel along the {\em same} direction in the $(x_{2},x_{1})$ plane. This new category of correlated interference, which is referred to as type II, is illustrated in fig. \ref{fig:barrier_negenergy} by the peak labeled $b$.  This is the same type of interference from reflections at the two surfaces of the slab modeled as two delta function in section \ref{sec:twodelta}. It is similar to the classical interference of a pulse of light reflecting from a thin film. 

Changing only the barrier spacing generates an oscillation in the PDF for peak $b$ analogous to that found in the interference of a pulse of light reflecting from a thin film whose thickness varies. That is, this peak goes through constructive and destructive interference from the two barrier reflections when the wavegroup size is much larger than that of the barrier and the spacing $D$ is varied. As time progresses peak $b$ maintains this interference as it travels in the $(x_{2},x_{1})$ plane. These are the same characteristics of the interference found in the simple two delta function model of reflection from a slab given in section \ref{sec:twodelta} and shown in fig. \ref{fig:slab}. This differs from the type I correlated interference associated with the fringes shown in the middle snapshot of fig. \ref{fig:barrier_negenergy} which is localized to a small temporal and spatial region.

This figure illustrates yet another form of correlated interference, referred to as type III: that from multiple reflections from the two barrier edges, which is shown as the fringes within the well of the middle snapshot of fig. \ref{fig:barrier_negenergy}. These are analogous to the standing wave formed in a thin film or optical cavity.

\paragraph{Particle traversing a finite barrier: $KE_{initial}^{relative} \approx PE$.}
\label{sec:barriermiddle}

Consider next wavegroups for which some Fourier components of the particle and barrier substates have a total initial kinetic energy in the relative coordinate system which exceeds the barrier potential energy while other components have a total relative initial kinetic energy which is less than the barrier potential. To illustrate the resulting PDFs the size of the particle and barrier substate wavegroups are chosen to be slightly larger than the barrier width $D$. 

\begin{center}
\begin{figure}
\includegraphics[scale=0.29]{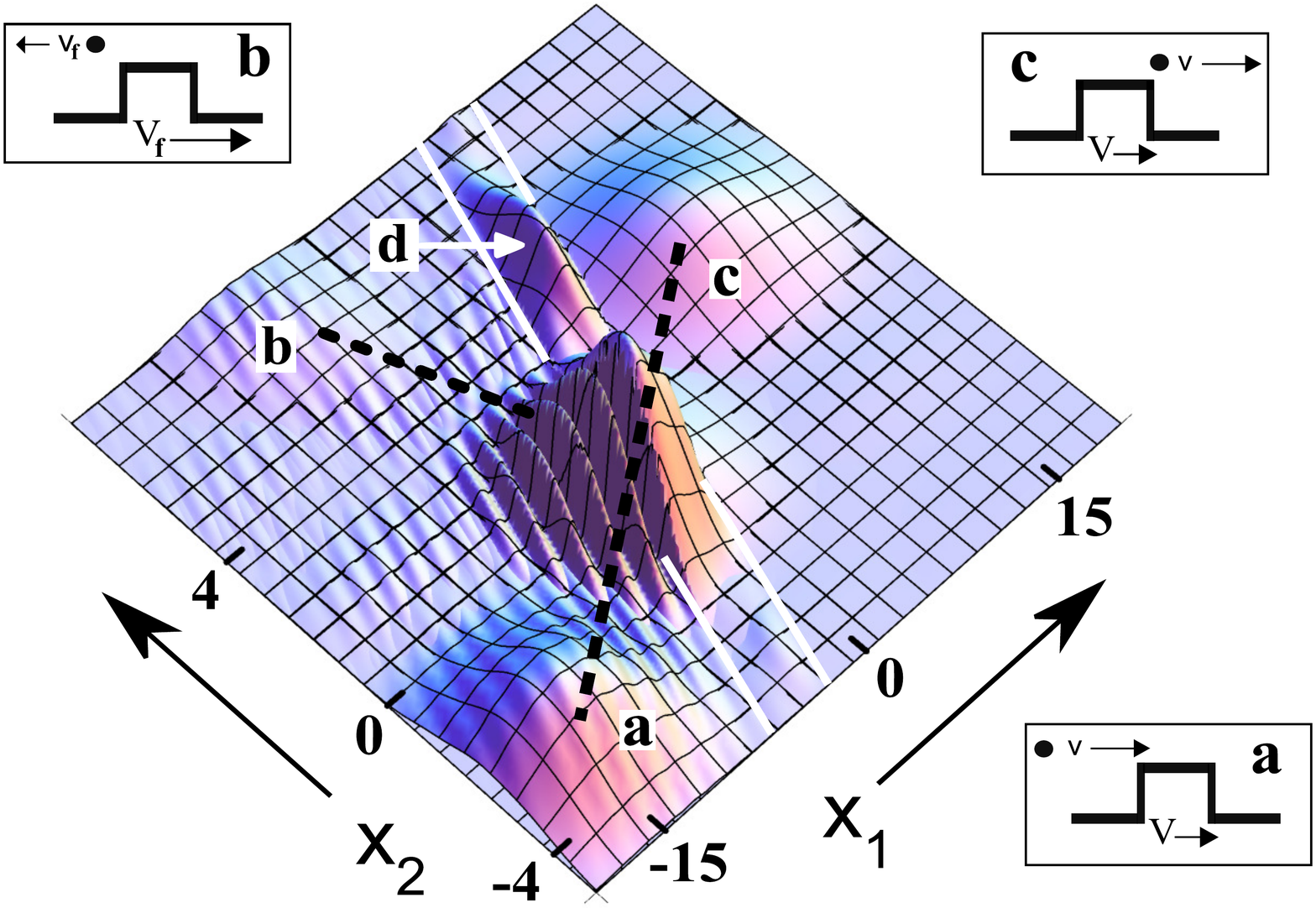}
\caption{PDF snapshots for three sequential times vs coordinates $(x_{2},x_{1})$ for the particle traversing the barrier. The only difference between the parameters used here and in fig. \ref{fig:barrier_negenergy} is the PE which forms a barrier.}
\label{fig:barrier_posenergy}
\end{figure}
\end{center}

Fig. \ref{fig:barrier_posenergy} shows the PDFs from such an interaction using three sequential snapshots, progressing along the dashed line from the lower left to upper right. The speed of the particle and well are illustrated for a classical system of two such particles in the insets next to each snapshot. Again the diagonal white lines correspond to $x_{1}=x_{2} \pm D$. The parameters used in fig. \ref{fig:barrier_posenergy} are $v_{0}/V_{0}=6$, $\Delta v/\Delta V=1.5$, $M/m=5$ while the $KE_{initial}^{relative}-PE/\mid PE\mid =0.3$ for the average value of the $KE_{initial}^{relative}$ for the wavegroup distributions. 

This figure also illustrates type III correlated interference but it differs from fig. \ref{fig:barrier_negenergy} in that only the first mode is excited. Additionally, the mode decays slowly enough to be visible in the third snapshot labeled as peak $d$ (located between both $x_{1}=x_{2} \pm D$ and peaks $b$ and $c$ in the third snapshot). This peak is analogous to the buildup and decay of electromagnetic energy in a optical cavity. Later snapshots (not shown) illustrate its decay. The peaks labeled $b$ in fig. \ref{fig:barrier_posenergy} and fig. \ref{fig:barrier_negenergy} are both of type II.

The position of peak $c$ can be compared between figs. \ref{fig:barrier_negenergy} and \ref{fig:barrier_posenergy} since all parameters are the same except the PE. The location of this peak indicates the effect of the interaction on the relative transit times for the particle and finite well or barrier wavegroup substates.

\subsubsection{Infinite well-particle wavegroup}
\label{sec:wellwavegroup}

The infinite well-particle calculation differs from that of the barrier due both to the particle and well velocities being constrained by the resonance condition given in eqn. \ref{eq:quantizationvV} and by the lack of coefficients, such as the $B$, $F$, $G$, and $H$ used in the previous section, which depend on the parameters of integration.

A closed form expression for the well substate wavegroup can be obtained from a Gaussian distribution of the energy eigenstates parameterized in terms of velocity components, $V$, of the well, given by
\begin{eqnarray}
\Psi_{well}^{wavegroup} \propto \int_{-\infty}^{\infty} \frac{e^{\frac{-(V-V_{0})^{2}}{2\Delta V^{2}}}}{\sqrt{\Delta V}} \Psi[x_{1},x_{2},t]dV  \notag
\label{eq:Psiwellwavegroup}
\end{eqnarray}
where the peak of the distribution is at $V_{0}$ and $\Delta V$ is its width. This is then summed over integral values of $n$ (the number of nodes) using the Gaussian distribution \cite{serge}
\begin{eqnarray}
\Psi_{total}^{wavegroup} \propto \sum_{n} \exp [-\{(n-n_{0})\pi \Delta x\}^{2}] 
\Psi_{well}^{wavegroup}, \notag 
\label{eq:Psiparticlewavegroup}
\end{eqnarray}
where the peak of the distribution is at $n_{0}$ and $\Delta x$ is its width.

\begin{center}
\begin{figure}
\includegraphics[scale=0.29]{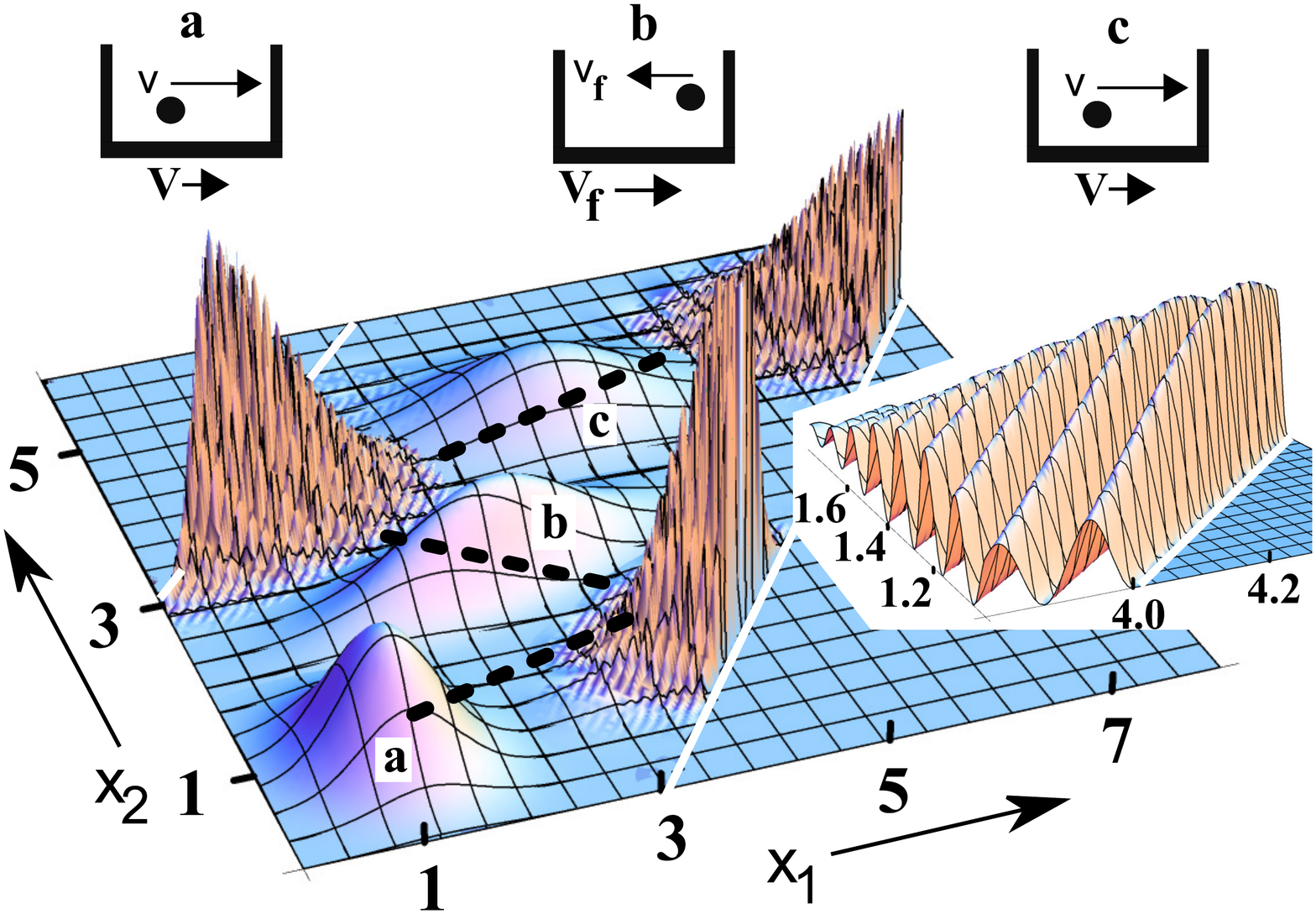}
\caption{PDF snapshots for six sequential times vs coordinates $(x_{2},x_{1})$ for a two-body wavefunction whose wavegroup size is less than the infinite well spacing. The diagonal white lines correspond to $x_{1}=x_{2} \pm D$. The trajectory is indicated by the dashed line. The right inset is a blow-up of the correlated interference of the second snapshot.}
\label{fig:wellsynchfinal}
\end{figure}
\end{center}

Using these relations, the PDF is plotted with particle and well substate wavegroups whose spatial widths are less than the well spacing $D$. Fig. \ref{fig:wellsynchfinal} shows such PDF results for six snapshots taken at equal time intervals progressing from the lower left to upper right along the dashed line. `Reflection' occurs at the diagonal white lines, corresponding to $x_{1}=x_{2} \pm D$.  The particle and well ``reflect'' from each other twice, in the second and fourth snapshots. The classical analogs for the particle and well positions for some snapshots are shown in the insets, labeled by a, b, and c. These correspond to the snapshots of the wavegroups labeled with the respective letters. While the insets are schematics of the `classical' analog between the wave and particle pictures, there is nothing similar for the correlated interference snapshots. Interference occurs when the incident and reflected two-body wavefunctions `overlap' and is shown in higher spatial resolution for the first reflection in the right side inset of fig. \ref{fig:wellsynchfinal}. The fringes are spaced by about half the deBroglie wavelength of the {\em particle} for $M>>m$ and $v>>V$, as are similar correlated interference fringes in two-body reflection of a particle from a mirror in fig. \ref{fig:interference}. The parameters used in fig. \ref{fig:wellsynchfinal} are $n_{0}=50$, $\Delta x/D=1/15$, $\Delta V/V_{0}=1/30$, and $M/m=10$. 

\section{Particle-only measurement to determine the mirror state}
\label{subsec:onlyparticle}

Correlation measurements are more difficult to perform than measuring just one of the bodies in a many body system. To predict one body effects, the two-body PDF is integrated (traced) over the coordinates of one of the bodies thereby converting it into a `one-body' or marginal PDF for the other body \cite{Gottfried}. 

For example, consider a measurement of the particle but not the mirror for the two-body PDFs shown in fig. \ref{fig:spreadvelocity}. Averaging over the mirror coordinate in fig \ref{fig:spreadvelocity}a essentially eliminates any interference when measuring only the particle. Averaging over the particle coordinate eliminates any interference in a measurement of only the mirror. Yet there is interference in a correlated measurement. Gottfried describes a related result using a different two-body interferometer \cite{Gottfried}.

Such averaging over the mirror coordinate of fig. \ref{fig:spreadvelocity}c, however, does not `wash out' interference in the particle-only measurement. The resulting particle-only `standing-wave' interference pattern in reflection from a static mirror is of course expected. Yet the integral of this two-body PDF over the particle coordinate yields no mirror-only interference. This matches the classically expectation that there is interference for the wave reflecting from a mirror but not for the mirror. However, if the coherence lengths are switched (it is difficult to increase the coherence length of a mesoscopic/macroscopic mirror) then a particle-only measurement will yield no interference while a mirror-only measurement will yield interference.

The two-body solution, when used in this manner to determine a one-body PDF, then reveals the limitations of starting with a solution to the one-body Schr\"odinger equation where the potential (mirror) is assumed to be a classical rather than a quantum object. For example, a one-body PDF for the particle, determined by the one-body Schr\"odinger equation, is a consequence of the solution to the two-body Schr\"odinger equation only for coherence lengths used in fig. \ref{fig:spreadvelocity}c. The reflected one-particle PDFs obtained when averaging over the mirror coordinates of fig. \ref{fig:spreadvelocity}a, \ref{fig:spreadvelocity}b, and \ref{fig:spreadvelocity}d, however, {\em cannot} be obtained without treating the mirror as a quantum object. The particle-only measurements, determined from this two-body treatment, then act as a probe of the mirrors quantum state.

\subsection{Particle reflection from a mirror}
\label{sec:particleonlymirror}

Two regimes in which the quantum nature of the mirror is revealed are illustrated in fig. \ref{fig:spreadvelocity}a and d using fixed particle bandwidth. The result in part \ref{fig:spreadvelocity}a was discussed above. The other regime, shown in fig. \ref{fig:spreadvelocity}d, requires such small mirror coherence lengths that the incident and reflected two-body states are distinguishable, thereby also eliminating any interference. 

Consider a simple estimate of the condition for which interference of the particle reflecting from the mirror disappears in the two-body PDF of fig. \ref{fig:spreadvelocity}d. This is done by approximating the incident two-body PDF of fig. \ref{fig:spreadvelocity}d as a rectangle with height $l_{c}^{mirror}$ and width $l_{c}^{particle}$ \cite{coherenceL}. Assuming a near stationary mirror with $m/M<<1$ and $V/v<<1$, reflection then tilts this rectangle to an angle $\beta$ given by $\tan \beta \approx V_{r}/v_{r}$ where $V_{r}$ and $v_{r}$ are the velocities after reflection of the mirror and particle. The overlap of the incident and reflected rectangles occurs at a distance $x$ from the point of maximum overlap given by $\tan \beta \approx l_{c}^{mirror}/x$. Beyond $x$ no overlap or interference occurs. Using the coherence length associated with thermal equilibrium of the mirror yields the following condition for the particle fringes to disappear in the two-body solution of fig. \ref{fig:spreadvelocity}d
\begin{equation}
l_{c}^{particle} \geq   \frac{h \sqrt{M}}{2m\sqrt{2k_{B}T}}.
\label{eq:SlabTemp3}
\end{equation}

This condition generates a regime in which the quantum nature of the mirror is manifest via elimination of the expected one-body interference of the particle in reflection, which is caused by the incident and reflected paths becoming distinguishable as the mirror coherence length becomes smaller (see section \ref{sec:bandwidth}). Although there is a lack of interference in the particle-only measurement, the two-body state is still split. This is evident in fig. \ref{fig:spreadvelocity}d as a bimodal distribution for which there is a finite probability of finding states of both reflection {\em and} no reflection.

The effect is enhanced with reflection of ultra-cold atoms which have both larger coherence lengths and masses than neutrons. A yet larger mass could involve a Bose-Einstein condensate itself if it reflects as ‘one particle’ rather than a collection of individual particles from the mirror \cite{condensate}. Such `standing-wave' interference on reflection of a Bose-Einstein condensate ($l_{c}^{condensate} \approx10^{-6}$m) is eliminated with a mirror of $M \leq 10^{-6}$ kg (the average mass of a mosquito) and $T=10$ K. Reflection of both atoms \cite{kouznetsov} and Bose-Einstein condensates \cite{pasquini} from mirrors has been studied but neither in the regimes discussed here nor in an attempt to measure such `standing-wave' interference.

\subsection{Particle reflection from a slab}
\label{sec:particleonlyslab}

Next consider measuring only the particle after reflection. For destructive interference of the joint PDF illustrated in fig. \ref{fig:slab} (or destructive interference of peaks b in figs. \ref{fig:barrier_posenergy} and \ref{fig:barrier_negenergy}), such averaging yields a low probability of measuring only the neutron even if the slab is {\em not} measured. In this case, destructive interference is maintained in the one-body neutron PDF.

The analogy between the measurement of a short pulse of light reflecting from a thin film and a measurement of only the neutron then has a similar physical interpretation. In both cases the detector need only be placed in the path of the moving neutron or the pulse of light, which leads to a one-body measurement that is independent of the wavegroup size. There is no requirement that the spatial resolution of the measuring instrument be smaller than the coherence length of the object being measured.

As in the mirror case, this example of destructive interference in a one-body measurement (which is derived from the two-body wavefunction) is valid only for a range of coherence lengths for both the slab and particle. Such coherence length dependence can be illustrated assuming a small reflection coefficient and using the mirror PDF of fig. \ref{fig:spreadvelocity}d for reflection from the first slab surface where the small slab coherence length is due to its mesoscopic/macroscopic mass. Consider a similar figure to represent the ``mirror'' reflection from the second slab surface but offset by a slab thickness, $D$, which is sufficient to prevent overlap of the reflected two-body wavefunctions from these two figures. This is a case where the particle does not experience interference even after having reflected from both slab surfaces and having a coherence length greater than the slab thickness. 

A simple estimate of this condition for the slab is given by again assuming that the incident two-body PDF of fig. \ref{fig:spreadvelocity}d is a rectangle of height $l_{c}^{mirror}$ and width $l_{c}^{particle}$. For a near sationary mirror with $m/M<<1$ and $V/v<<1$, the rectangles reflected from each surface are tilted at an angle $\beta$ given by $\tan \beta \approx V_{r}/v_{r}$. The positions of the slab surfaces are essentially vertical in the $x_{1},x_{2}$ plane. The two reflected rectangles just separate when $\tan \beta \approx l_{c}^{mirror}/D$.  Using the coherence length associated with thermal equilibrium of the slab with its environment yields the temperature condition for which there is no interference
\begin{equation}
T>h^{2}M/(8D^{2}k_{B}m^{2}).
\label{eq:SlabTemp2}
\end{equation}
Note the similarity with eqn. \ref{eq:SlabTemp} where the condition on $T$ is for interference rather than for lack thereof.

Such elimination of interference is due to the two particle-slab reflected states becoming distinguishable as the slab coherence length decreases in a manner similar to that discussed for the particle-mirror system in section \ref{sec:bandwidth} and is only a consequence of treating both the particle and slab as quantum objects. If the slab is not treated as a quantum object then particle interference will always occur as long as the particle coherence length is greater than the slab thickness. A similar effect is found in the particle-well model.

Consider now the opposite extreme in the quantum character of the slab, that of small slab bandwidth or large uncertainty in the slab position. Unlike the case of large mirror uncertainty where averaging over the mirror position in fig. \ref{fig:spreadvelocity}a `washes out' interference, uncertainty in the slab position does not eliminate the one-body-only measurement of the interference for the particle reflecting from both slab surfaces. This is a consequence of that interference not depending on the slab location.

The inequality in eqn. \ref{eq:SlabTemp2} is most easily satisfied by increasing the mass of the reflected particle (increasing the recoil of the slab) and/or that of the slabs spacing (requiring an increase in the coherence length of the particle) while decreasing the mass of the reflector (increasing the slab recoil). Reflection of a neutron traveling at $v=1000$m/s from a slab at $T=100$K, as in the example used above, yields no interference for $M \leq 10^{-24}$ kg. Elimination of `thin-film' interference for a mesoscopic/macroscopic slab requires larger $m$ and $D$ (with the size of the mesoscopic/macroscopic mass decreasing quadratically with these parameters).

\section{Decoherence}
\label{sec:decoher}

Interference in a typical one-body interferometer `washes-out' as the fringe spacing becomes imperceptible with increasing mass. Another mechanism that eliminates interference is a measurement to determine along which path the object traveled through the interferometer. The former issue was mitigated using two-body reflection of a microscopic particle from a mesoscopic/macroscopic object, as described above, while the latter is the first topic of this section.

\subsection{Path information}
\label{sec:path}

A method to determine path information and therefore destroy interference utilizes a `Heisenberg microscope,' where an additional particle scatters from the object traversing the interferometer. If the wavelength of this particle, $\lambda$, is smaller than the spatial separation along the two paths, $\Delta x$, then path information can be obtained in principle from the state of this scattered probe particle and interference is destroyed. This was verified using a three grating Mach-Zehnder interferometer traversed by an atom and using a photon as the probe to determine path information \cite{chapman}. The contrast of the interference, as a function of $\Delta x$, was shown to drop dramatically for $\Delta x>\lambda/2$.

Generating such decoherence for particle-mirror interference involves using a probe particle to determine the mirrors position during interference. The `two paths' exist only during the time that the incident and reflected two-body wavegroup states overlap for reflection from a mirror. During this time there are two distances that the mirror could have moved: that associated with and that without reflection. If the probe particle can resolve the difference in these distances then the `path' associated with reflection is distinguishable from that without reflection. For a near stationary mirror with $m/M<<1$ and $V/v<<1$, the mean difference in these mirror distances is $\Delta x\approx 2l_{c}^{particle}m/M$, which when equal to the wavelength of a probe particle of mass $m^{*}$ then requires that the probe have velocity $v_{probe}\approx hM/(4l_{c}^{particle}mm^{*})$.

This calculation can also be used to approximate the effect of such decoherence on the mirror from the collisions with gas atoms forming the thermal bath which surrounds the mirror if these gas atoms are considered to be the probe which determines path information and they scatter from only one of the superposed mirror states. The interference in reflection described above is then maintained if there is a small probability of these atoms having the velocity, $v_{probe}$. For a mesoscopic/macroscopic mirror and both a microscopic particle and a probe, $\Delta x$ is too small to be resolved by typical low temperature gas atoms.

Next this decoherence mechanism is applied to slab reflection. The two surface reflections cause a difference in slab displacement which was given in section \ref{sec:twodelta} for a near sationary mirror with $m/M<<1$ and $V/v<<1$, as $\Delta\approx 2mD/M$. Determining from which slab surface the particle reflects then requires that the probe have velocity $v_{probe}\approx hM/(4Dmm^{*})$.

This decoherence mechanism can also be applied to the emission of photons by the slab between the times of particle reflection from each slab surface. If the photons wavelength is small enough to allow for the location of the slab associated with one of the reflections ($3 \times 10^{-22}$ m in the neutron-slab example given above), interference is destroyed since it can then be determined at which slab surface the reflection occurred \cite{cronin}. Obviously the probability of such a thermal photon emission is small.

These calculations indicate that environmental decoherence associated with obtaining path information can potentially be mitigated. This is due in part to the displacements associated with the interfering reflector substates being proportional to $m/M$. To measure such small displacements requires scattering particles with proportionally smaller wavelengths which are typically not found in the environment.

\subsection{Random phase}
\label{sec:randomphase}

It should be pointed out that the calculations presented above assume that states for the particle, mirror, and slab have a minimum spatial profile (minimum uncertainty Gaussian wavepackets). This will be referred to as a phased state. Introducing a random phase into the Fourier components of either one or both of the substates of such an incident two-body phased state affects the shape and size of the wavepacket while maintaining the same bandwidth and therefore the same coherence length. This is referred to as a dephased state. Such random phase structure is introduced as an initial condition but could have been imposed via an earlier measurement which collapsed the reflector state into a dephased state.

Consider an incident phased particle substate and incident dephased mirror substate which then reflect. A one-body particle measurement after reflection can then act as a sensitive probe of this mirror substate. To illustrate this consider how the PDF of fig. \ref{fig:spreadvelocity}c changes for such a dephased state. First, the PDF is spread spatially along the $x_{2}$ axis. Second, the magnitude of the interference part of the PDF is greatly reduced but not eliminated.  Finally, the PDF is not spatially uniform. Apart from its magnitude and spatial uniformity, this PDF is more similar to fig. \ref{fig:spreadvelocity}a than \ref{fig:spreadvelocity}c. This is not surprising since the localization of the dephased mirror substate is similar to that of the mirror in fig. \ref{fig:spreadvelocity}a.

Results of a particle-only measurement are predicted from an integral (trace) over the mirror coordinate of this PDF which then `washes-out' this interference. If the mirror is in a dephased state, a reflecting particle-only measurement will not exhibit interference in the overlap region. However, such interference is expected both when treating the mirror as a classical potential and for a phased mirror substate with the coherence lengths similar to those shown in fig. \ref{fig:spreadvelocity}c. 

The spatial extent of such a dephased mirror substate can vary between being unlocalized to being the minimum size determined by the bandwidth. This size can then be probed by varying the wavelength of the incident particle in a reflected particle-only measurement, in which case interference either disappears or reappears as this wavelength becomes smaller or larger than the spatial size of the dephased mirror substate.

\subsection{Environmental decoherence}
\label{sec:environ}

Yet another decoherence mechanism is through interaction with the environment. Zurek predicted exponential decay of the off-diagonal density matrix element terms for a macroscopic object in a superposition state, $\exp[-t/t_{D}]$, with time constant $t_{D}=t_{R}(\lambda_{T}/\Delta x)^{2}$ where $t_{R}$ is the thermal relaxation time, $\Delta x$ is the difference in distance between these two states, and $\lambda_{T}$ is the thermal de Broglie wavelength of the macroscopic object \cite{zurek}. The decay is into a mixed rather than a pure state. It is interesting to note that the lack of interference predicted in section \ref{subsec:onlyparticle} also depends on $\Delta x$ but for very different reasons. 

Applying this decoherence model to the macroscopic slab states after reflection of the neutron, as described above, yields $t_{D}/t_{R}\approx M h^{2}/(8k_{B}TD^{2}m^{2})$. For these neutron parameters this ratio is large ($\approx5\times10^{14}$ for $M=10^{-8}$kg at $300$K) due to the very small offset introduced from the recoil ($\lambda_{T}>>\Delta x$). The parameter of interest, $t_{D}$, is also affected by $t_{R}$ which can vary dramatically depending on environmental and material properties along with the dimensional values of the slab. Mesoscopic masses can have short thermal relaxation times. Nevertheless, it appears that environmental decoherence for reflection from a slab can easily be made negligible in this model of decoherence. 

Applying this decoherence model to the mesoscopic/macroscopic mirror states during correlated interference in reflection allows a greater variation of parameters since atoms or molecules can more easily reflect from a surface than from both surfaces of a `thin film.' The result, $t_{D}/t_{R}\approx M h^{2}/(8k_{B}T(l_{c}^{particle}m)^{2})$, therefore allows environmental decoherence for the mirror to more easily be made either negligible or non-negligible.

Consider next the consequences of such decoherence on correlated interference where the particle and mirror are in superposition states of both having and not having reflected. If this mirror substate decoheres into a mixed state then it seems reasonable to assume that the commensurate particle substate must also decohere into a mixed state since the two substates are correlated via conservation of energy and momentum. Measurement of only the particle, after the mirror has decohered, therefore yields no interference since the particle cannot then be in a superposition state.

Applying this result to a measurement of only the neutron flux reflecting from the slab, while varying the incident neutron's wavelength, results in no interference if the slab decoheres during the time between reflection from both surfaces. The details of actually executing such a measurement and of choosing parameters which straddle the region where decoherence is and is not applicable are of course much more complicated and are beyond the scope of this work.

Braginsky and Khalili also treat environmental decoherence  \cite{braginsky}. However, they predict that for a macroscopic oscillator to exhibit quantum effects both the mechanical and thermal relaxation times are important while the parameter $\Delta x$ is not incorporated in their model. They argue that quantum effects can indeed be exhibited for macroscopic high Q resonators.

\subsection{Decoherence overview}
\label{sec:overview}

It has been shown that correlated interference in reflection does not disappear with increasing mirror mass, is difficult to eliminate via a path measurement, and can be made insensitive to environmental decoherence. It should also be pointed out that the robust character of interference for objects with many degrees of freedom is reinforced by measurements which demonstrate that even if the size of the object is larger than both the coherence length and deBroglie wavelength, interference can still be observed \cite{cronin}.

\section{Discussion and Summary}
\label{sec:summary}

The use of reflection to generate perceptible interference in a mesoscopic/macroscopic object involves a microscopic particle interacting with a mesoscopic/macroscopic reflector to create two reflector substates which differ only by the microscopic momentum exchanged. Unlike the fringes for a particle traversing a double slit, the interference fringes for the reflector substates do not vanish in the limit of large reflector mass. 

In addition, particle and reflector interference is required for conservation of energy and momentum to be satisfied. For example, consider reflection of a neutron from a slab. `Thin-film' interference yields peak b in the two-body PDF of fig. \ref{fig:barrier_negenergy}. Destructive interference of the reflected neutron substate, which is certainly expected in a one-body treatment, must be associated with destructive interference of the slab substate (peak b must not exist). That is, if the neutron can not be measured in reflection then the slab can never be measured at the position associated with it either having or not having reflected the neutron. Similarly, by only changing the neutron wavelength to generate constructive 'thin-film' interference in reflection, the slab must then have a commensurate probability to be found in a position of having reflected the neutron. Measurement of interference in the reflected particle then requires that the reflector be in a superposition state which interferes in a manner correlated with that of the particle. This conclusion also holds for the interference shown in fig. \ref{fig:interference} and presumably for all two-body interferometers.

Existence of such interference might be disputed since the difference in wavevectors of the interfering mirror substates, due to reflecting a microscopic particle, is much smaller than the spread in wavevectors of which the mirror wavegroup is composed, due to its thermal motion. This can be expressed as the ratio $R=\frac{\delta K_{reflection}}{\delta K_{thermal}}$. Interference, however, is not determined by $R$ but rather by the relation between the interferometer path difference and the coherence lengths of the interfering wavegroups.  For a neutron reflecting from a mirror of $M=10^{-10}$ kg, $v=100$ m/s, and $T=10$ K, the ratio $R \approx m v/\sqrt{Mk_{B}T}$ yields a number which is of the same order of magnitude as is this ratio for an optical pulse of duration $10^{-13}$ s (from a mode locked Ti-Sapphire laser) traversing a Mach-Zehnder interferometer, with the mirror in one arm moving at speed $10^{-2}$ m/s (all other interferometer components are static).  Optical pulse trains exhibit interference in similar interferometers \cite{lepetit} with smaller $R$ values than in the particle reflection example just given. 

Correlated interference in reflection is an example of one of the simplest interferometers, utilizing neither division of amplitude nor division of wavefront methods to generate interference. In addition, path lengths need not be carefully matched for interference to be manifest.

The complexity of this interferometer is revealed as the coherence lengths of the particle and mirror substates are varied. Examples are: (1) When both are small, classical reflection rather than interferometry essentially occurs. (2) When both are large, compared with the particle's wavelength, reflection becomes an interferometer yielding correlated interference, as shown in fig. \ref{fig:spreadvelocity}a, while no interference is observed in a one-body-only measurement of either the particle or mirror. Due to uncertainty in the location of both objects, a measurement of the particle at $x_{1}$ then interferometrically correlates with a range of position measurements of the mirror. (3) When the mirror's displacement, due to recoil in reflection, is less than the mirrors coherence length and the coherence length of the particle is larger than that of the mirror, then a measurement of only the particle yields the expected interference from a one-body Schr\"odinger equation treatment (an example is shown in fig. \ref{fig:spreadvelocity}c). (4) When the coherence length of the particle is much larger than that of the mirror and mirror recoil is larger than its coherence length, reflection becomes a two-body beamsplitter. This is illustrated in fig. \ref{fig:spreadvelocity}d for $x_{1}<-100$. Here two states are generated but do not interfere, one being the particle-mirror state before reflection and the other that after reflection, similar to a beamsplitter producing photon states which traverse spatially separate and therefore distinguishable paths. For example, at fixed $x_{1}$ there is a bimodal PDF along the $x_{2}$ axis, each peak of which corresponds to one of these two states. There is no interference since the short coherence length of the mirror relative to the recoil distance results in distinguishable two-body states (the mirror has reflected the particle in the upper but not the lower PDF peak).

The kinematics in the correlated interference regime can be compared with that of an electromagnetic wavegroup reflecting classically from a mirror. The latter involves accelerated motion due to a continuous force delivered to the mirror throughout the reflection process. To determine the former, consider the interference region shown in fig. \ref{fig:interference} with $v/V=10$, $m/M=1$, and equal coherence lengths. If only the mirror is measured (integrating the PDF over the particle coordinate) then the expectation value for the mirrors position accelerates throughout the interaction time, as expected classically. If the mirror is not measured but the particle is, then the expectation value for the mirrors position decelerates at the same rate. However, these expectation value kinematics vary with parameters such as particle and mirror substate coherence lengths and masses. 

Finally, the issue of how to determine that the mirror is in a superposition of states is considered. One method is to perform a correlation measurement of the positions of the particle and cm of the mirror in the interference region. In a region of destructive interference neither will be found. However, correlation measurements are difficult to perform.

Another possibility is to use a third particle as a probe, interacting only with the mirror to determine if the mirror is in a superposition state, in the following way: let this microscopic probe particle reflect from both mirror states (reflecting from the other side of the mirror) resulting in one-body interference of the probe. The coherence length of the probe need only be larger than the very small displacement of the mirror states for this probe interference to be exhibited, while the magnitude of the phase shift is determined by the mass of the probe. This is a method to measure that the mesoscopic/macroscopic mirror is in a superposition state, without any direct measurement of the mirror. However, calculations of such three-body interactions, which follow easily from the formalism discussed above, are beyond the scope of this already lengthy foundational work.

Methods to measure quantum effects on a mesoscopic/macroscopic reflector using only the reflected particle have been a focus of this work, rather than the methods just described. The reflected particle flux is then used to reveal properties of the quantum state of the reflector. One example is of a neutron reflecting from a slab. As with interference in reflection of light, the one-body Schr\"odinger equation predicts a variation in reflected neutron flux as a function of incident neutron wavelength whenever the neutron coherence length is longer than the slab thickness. 

The calculation presented here, treating both the particle and reflector as quantum objects, also predicts such interference but only as a special case. In addition, this treatment predicts the following regimes under which no reflected interference occurs {\em for a measurement of only the particle}: (1) The reflection process acts as a beamsplitter and not an interferometer, generating two mirror or slab states which do not overlap. (2) The Fourier components of the mirror substate have random phases sufficient to extend its size beyond that of the particle's wavelength. (3) The slab and mirror have coherence lengths long enough for correlated interference but not interference when only the particle is measured. 

In each of these cases, consider varying the incident particle's wavelength and observing no interference in the reflected particle flux. This indicates that the mesoscopic/macroscopic reflector is in a particular quantum superposition state. By varying parameters appropriately, as discussed in section \ref{sec:particleonlymirror}, interference in this reflected particle flux then reappears. Ironically, it is this lack of interference that then reveals the quantum nature of this mesoscopic/macroscopic `thin film' or mirror. In this case, measuring only the reflected particle flux yields significant experimental advantages over a measurement of the correlation between the particle and mirror.

Another possible method to study the state of the mesoscopic/macroscopic mirror is via the expected standing wave pattern generated between incident and retro-reflected states of a Bose-Einstein condensate as it interacts with a mirror. If this process can be modeled as described above, with the mass of the condensate replacing that of the particle, then the point at which condensate interference disappears can be extended to larger mirror masses. Although interference between split condensates has been observed {\cite {kohstall,witkowski}}, as far as we are aware, there is no evidence of standing wave interference due to reflection.

Reflection of atoms from mirrors has been previously studied but not in treating both the particle and reflector as quantum objects nor in the reflector mass regimes discussed here nor in an attempt to measure either the resulting `standing-wave' interference or the effect of decoherence on such interference. There exist few examples of quantum correlated interference for non-zero rest mass particles. Verification of the above predictions, even in the microscopic regime for both particle and reflector, is therefore of fundamental interest.   Although far from being comprehensive, these results indicate a direction, heretofore unexplored, for further research in understanding quantum correlation, probing decoherence, and extending quantum measurements to larger masses.

\begin{acknowledgments}
The authors would like to thank Professors C. Durfee and J. Scales for useful comments.
\end{acknowledgments}

% ----------------------------------------------------------------

\begin{references}
\bibitem{currentdensity} In the particle-mirror coordinates the incident and reflected wavevectors for the particle (or mirror) are not necessarily equal and opposite as they are in the cm/rel coordinates (nor are they equal and opposite for a one particle system in which the mirror is fixed). The probability current density to the left of the mirror then exhibits spatial and temporal interference due to the effect this difference in wavevectors has on the incident and reflected wavefunctions. This results in a non-additive current density, a consequence of the space to the left of the mirror increasing as it moves. The net flux  must account for the build-up of probability within this ``new'' space.
\bibitem{coherenceL} The coherence length is used here rather than the wavepacket size since the former rather than the later determines interference, op. cit. \cite{Hamilton}.
\bibitem{condensate} If the condensate particles reflect individually from a receding mirror then they each acquire a different energy which decreases in their order of reflection. This is due to the energy exchanged with the mirror from the particles which have already reflected and is manifest as a distorted reflected wavegroup.
\end{references}
\end{document}